\documentclass[11pt,a4paper]{article}
\usepackage{amsfonts,amssymb,amsmath,amsthm,cite}
\usepackage{graphicx}

\evensidemargin=\oddsidemargin
\advance\topmargin by -\headheight
\advance\topmargin by -\headsep

\newtheorem{assumption}{Assumption}[section]
\newtheorem{theorem}[assumption]{Theorem}
\newtheorem{corollary}[assumption]{Corollary}

\newtheorem{lemma}[assumption]{Lemma}
\newtheorem{definition}[assumption]{Definition}
\newtheorem{remark}[assumption]{Remark}
\newcommand{\Sp}{{\rm Sp}}
\newcommand{\Tr}{{\rm Tr}}
\newcommand{\Th}{\Theta}
\renewcommand{\th}{\theta}
\newcommand{\A}{\mathcal{A}}            %% an algebra
\renewcommand{\a}{\alpha}               %% short for \alpha
\newcommand{\C}{\mathbb{C}}             %% complex numbers
\newcommand{\Coo}{C^\infty}             %% space of smooth functions
\newcommand{\DD}{\mathcal{D}}           %% Dirac operator
\newcommand{\eps}{\epsilon}             %% short for \epsilon
\newcommand{\Ga}{\Gamma}                %% short for \Gamma
\newcommand{\ga}{\gamma}                 %%short for \gamma
\renewcommand{\H}{\mathcal{H}}          %% Hilbert space

\newcommand{\Isom}{{\rm Isom}}
\renewcommand{\L}{\mathcal{L}}          %% operator algebra
\newcommand{\Mop}{\star}                %% Moyal product (just *)
\newcommand{\N}{\mathbb{N}}             %% nonnegative integers
\newcommand{\ox}{\otimes}               %% Short for tensor product 
\newcommand{\pa}{\partial}
\newcommand{\R}{\mathbb{R}}             %% real numbers
\newcommand{\rad}{{\mathbf r}}
\newcommand{\sepword}[1]{\quad\mbox{#1}\quad} %% well-spaced words
\newcommand{\set}[1]{\{\,#1\,\}}        %% set notation
\renewcommand{\SS}{\mathcal{S}}         %% Schwartz space
\newcommand{\Sf}{\mathbb{S}}
\newcommand{\T}{\mathbb{T}}             %% circle as a group
\newcommand{\thalf}{\tfrac{1}{2}}       %% small* fraction  1/2
\newcommand{\tri}{\Delta}               %%short for Laplacian
\newcommand{\vf}{\varphi}               %% short for \varphi 
\newcommand{\x}{\times}                 %% cartesian product or cross
\newcommand{\Z}{\mathbb{Z}}             %% integers
\newcommand{\diag}{{\rm diag}}
   
\DeclareMathOperator{\Mod}{mod}

\begin{document}

\thispagestyle{empty}
\quad

\vspace{2cm}
\begin{center}

{\huge
\textbf{Heat kernel and number theory on NC-torus}}

\vspace{1.5cm}

{\large V. Gayral$^1$, B. Iochum$^{2,3}$ and D.~V. Vassilevich$^{4,5}$} \\

\vspace{1.5cm}

{\large\textbf{Abstract}}

\end{center}

\begin{quote}
The heat trace asymptotics on the noncommutative torus, where generalized
Laplacians are made out of left and right regular representations, 
is fully determined. It turns out that this question is very sensitive
to the number-theoretical aspect of the deformation parameters. The
central condition we use is of a Diophantine type. 
More generally, the importance of number theory is made explicit 
on a few examples. We apply the results to the spectral action computation and
revisit the UV/IR mixing phenomenon for a scalar theory. 
Although we find non-local
counterterms in the NC $\phi^4$ theory on $\T^4$, we show that
this theory can be made renormalizable at least at one loop, and
may be even beyond.
\end{quote}

\vspace{1cm}

July 2006

\vspace{1cm}

%%%% NCFT;
%%%% mathematical methods in physics: functional analysis;
%%%% classical and semiclassical techniques in gauge field theories,
%%%% respectively

%%%% Topological algebras of operators;
%%%% Noncommutative function spaces;
%%%% NCG \`a la Connes;
%%%% Phase space methods including Wigner distributions,
%%%% respectively

\vspace{2cm}

\noindent 
$^1$ Matematisk Afdeling, K\o benhavns Universitet, Denmark, 
gayral@math.ku.dk,\\
$^2$ UMR 6207

-- Unit\'e Mixte de Recherche du CNRS et des
Universit\'es Aix-Marseille I, Aix-Marseille II et de l'Universit\'e
du Sud Toulon-Var

-- Laboratoire affili\'e \`a la FRUMAM -- FR 2291\\
$^3$ Also at Universit\'e de Provence, iochum@cpt.univ-mrs.fr,\\
$^4$ Institut f\"ur Theoretische Physik, Universit\"at Leipzig, 
Germany, Dmitri.Vassilevich@itp.uni-leipzig.de\\
$^5$ V.~A.~Fock Institute of Physics, St.~Petersburg University, Russia

\newpage

\section{Introduction}

The importance of heat kernel techniques in spectral analysis (see 
\cite{newGilkey,Kirsten}) or in quantum field theory is known for a long time 
(see for instance the references in \cite{Vassilevich:2003xt}). This
type of expansion is particularly very useful for the control of 
anomalies and loop divergences. Naturally, its extension to 
noncommutative theories using for instance the Moyal product instead 
of the pointwise one, has also begun a long time ago (see reviews 
\cite{Douglas, Szabo, Varilly} and also \cite{CDS}). 
The idea, originally due to Heisenberg, 
behind this generalization is that it could help to suppress 
some divergences. Unfortunately, a 
consequence of this idea is that the situation is as 
difficult as in the classical setting, or even worse since some 
UV/IR mixing can occur, except in some peculiar cases 
where the renormalisability of the model is proved 
\cite{Grosse:2004yu}. Meanwhile, the noncommutative geometry (NCG) 
pioneered by Alain Connes \cite{Book} has shown its capacity to cover 
isospectral deformations like the deformation of a classical torus into 
the celebrated noncommutative torus (nc-torus). While many physical 
ideas coming from string theory have justified a systematic study of 
noncommutative quantum field theory, the interest of NCG stems also from 
its mathematical roots. In particular, the spectral action introduced 
by Chamseddine--Connes refers to a spectral triple $(\A, \H,\DD)$ of 
an algebra $\A$ acting on a Hilbert space $\H$ and a given Dirac 
operator $\DD$ which generates the inner fluctuations corresponding 
to gauge potentials. This spectral action is simply $\Tr\big( 
\Phi (\DD_{A}/\Lambda)\big)$ where $\Phi$ is a positive even 
function, $\Lambda$ is a mass scale parameter, $\DD_{A}=\DD + A$ 
and $A$ is a one-form.

This torus depends on parameters through a deformation matrix $\Th$ and it 
appears that the heat asymptotics are very sensitive to it. In particular,
some Diophantine conditions, see (\ref{Diophcond}),
(\ref{Diop}), are necessary to control the number-theoretic deviation 
from rational numbers. We also investigate a situation beyond this 
condition which yields precise aspects of number theory.

To control the heat trace asymptotic we apply a two-step
procedure. First, we define a trace (cf. (\ref{defSp})), which is
indeed proportional to the Dixmier trace, and calculate it through
the Fourier coefficients (cf. (\ref{eSp})). Then we prove that
being expressed in terms of this trace the heat trace asymptotics
for generalized Laplacians look precisely the same as in the
commutative case.

We first apply the control of the heat trace asymptotic to the
spectral action computation.
This action has been {\it partially} computed in 
\cite{GI2002}, but here we pay attention to the natural existing 
real structure $J$ of 
the triple (\cite{ConnesReality}): now 
$\DD_{A}=\DD + A+ \epsilon JAJ^{-1}$, 
so we are in the most difficult situation where exist simultaneously
left and right regular representations (see also \cite{GBIS} for 
further physical motivations). 
We show here that this {\it full} spectral action is the expected one
for a 4-dimensional nc-torus. The amazing fact is that unlike for 
arbitrary generalized Laplacians, where non-standard terms appear in
the heat kernel expansion (typically of the form `product of traces'), for
the square of the covariant Dirac operator, such weird terms are absent.
Thus the formula (\ref{spectralaction1}) we obtain is the expected
one, up to some numerical
coefficients.  

Then, we apply our results to the study of a 
scalar field in an nc-4-torus. We show that the divergent part of 
the effective action does not reproduce the structure of the classical 
one, that is, the divergences cannot be cancelled by a proper 
couplings re-definition. Nevertheless, the theory can be made renormalizable 
at one loop by adding to the classical action a non-local term, which
perfectly fits in with the philosophy of \cite{Grosse:2004yu}. We
conjecture that the modified theory is renormalizable to all orders in
perturbation theory.

The paper is organized as follows: we recall in Section 2 some useful 
facts on nc-tori and study a trace, in fact a Dixmier trace, applied 
to operators like $L(a)R(b), \, a,b\in \A$, 
where $L$ (resp. $R$) is the 
left (resp. right) multiplication, giving a full asymptotic of  
$\Tr\big(L(a)R(b)e^{-tP}\big)$ as $t \to 0$ for a 
generalized Laplacian $P$. Section 3 touches on toric noncommutative 
manifolds, not necessarily compact. The spectral action is computed 
in Section 4 and the last section is devoted to the study of
divergences of a scalar field theory.

Since the proof of the asymptotics of the heat trace is technical, it 
is postponed to Appendix A while some consequences of number theory 
in this setting are developed in Appendix B.

\section{Heat trace asymptotic on NC-torus}
\subsection{Traces and number theory}

Let $\Coo(\T^n_\Th)$ be the smooth noncommutative 
$n$-torus associated to a skewsymmetric deformation matrix $\Th \in 
M_n(\R)$ (see \cite{ConnesTorus}, \cite{RieffelRot}). This means that 
$\Coo(\T^n_\Th)$ is the algebra generated by $n$  
unitaries $u_i$, $i=1,\dots,n$ subject to the relations
\begin{equation}
\label{rel}
u_i\,u_j=e^{i\Th_{ij}}\,u_j\,u_i,
\end{equation}
and with Schwartz coefficients:
using the Weyl elements $U_k:=e^{-\frac i2 k.\chi
k}\,u_1^{k_1}\cdots 
u_n^{k_n}$, $k\in\Z^n$, relation
\eqref{rel} reads 
\begin{equation}
\label{rel1}
U_{k}U_{q}=e^{-\frac i2 k.\Theta q} \,U_{k+q} \,,
\end{equation}
where $\chi$ is 
the matrix restriction of $\Theta$ to its upper triangular part. 
Thus unitary operators $U_{k}$ satisfy $U_{k}^*=U_{-k}$ and a typical 
element $a\in\Coo(\T_\Th^n)$ can be written as 
$a=(2\pi)^{-n/2}\sum_{k\in\Z^n}a_k\,U_k$, where $\{a_k\}\in\SS(\Z^n)$.
We use this non-standard normalization in order to simplify upcoming
formulas. 

Let $\tau$ be the (unique) normalized
faithful trace on $\Coo(\T^n_\Th)$ defined by 
$$
\tau\big(a\big):=(2\pi)^{-n/2}a_0
$$ 
and $\H_{\tau}$ be the GNS Hilbert space obtained 
by completion of $ \Coo(\T_\Th^n)$
with respect of the norm induced by the scalar product 
$\langle a,b\rangle:=\tau(a^*b)$.

On $\H_{\tau}$, we consider the left and right regular
representations of
$\Coo(\T_\Th^n)$ by bounded operators, that we denote respectively 
by $L(.)$ and $R(.)$.

An easy consequence of the associativity of the algebra 
is the commutativity of these two representations,
namely $L(a)R(b)=R(b)L(a)$, for all $a,b\in\Coo(\T_\Th^n)$.

Let also $\delta_\mu$, $\mu=1,\dots,n$ be the $n$ (pairwise
commuting)
canonical derivations, defined by
\begin{equation}
\delta_\mu(U_k):=ik_\mu U_k. \label{dUk}
\end{equation}
They extend to unbounded operators on $\H_\tau$
(with suitable domain),
and let $$\tri:=- g^{\mu\nu}\,\delta_{\mu}\delta_{\nu} \geq 0$$ 
be the associated Laplacian where we assume that the metric $g^{\mu 
\nu}$ on $\T^n$ is constant.

In the following, for $k,q \in \Z^n$, we denote  
$k.q:=g_{\mu\nu}k^{\mu}q^{\nu}$ and $\vert k \vert^2:=k.k$. 
Naturally, the matrix $\Theta$ satisfies 
$k.\Theta q=-\Theta k.q$.

\quad

There is (at least) one analogous description of $\Coo(\T_\Th^n)$
given in
terms of Rieffel star-product \cite{ReiffelDefQ}. If $\alpha$ denotes
the
(periodic) action of $\R^n$ on $\T^n$, then $\Coo(\T_\Th^n)\simeq
\big(\Coo(\T^n), \Mop\big)$, where the star-product $\Mop$ is defined
by
the following oscillatory integrals:
\begin{equation}
\label{Rieffel}
f\Mop h:=(2\pi)^{-n}\int_{\R^n\times\R^n} 
d^ny\,d^nz\,e^{-iyz}\,\alpha_{\thalf\Th
    y}\big(f\big)\, \alpha_{-z}\big(h\big),\quad f,g\in\Coo(\T^n),
\end{equation}
yielding relation \eqref{rel1} on the Fourier modes,
\begin{equation*}
e^{ikx} \star e^{iqx}=e^{-\frac i2 k.\Theta q} e^{i(k+q)x}.
\end{equation*}
A counterpart of eq.\ (\ref{dUk}) reads 
$\partial_\mu e^{ikx}=ik_\mu e^{ikx}$. In this
description, the above defined Laplacian is nothing else but the 
ordinary one
(associated to the constant metric $g$) and the trace $\tau$ is the
normalized integral, 
\begin{equation*}
\tau (f):=(2\pi)^{-n} \int_{\T^n} d^nx\, f(x) =
({\rm vol}\, \T^n)^{-1}\int_{\T^n} d^nx\, \sqrt{g}\, f(x),\quad f\in\Coo(\T^n).
\end{equation*} 
We can consider using a
non-flat metric, but we need (for later use) a severe 
restriction on it: the $\R^n$-action must be isometric. Thus only
constant metrics are allowed.\footnote{There are very few attempts to
deal with metrics which are not constant in the non-commutative directions.
So far, one was able to obtain expressions for the heat trace asymptotics
as formal power series in deviations of the metric from the flat one only
\cite{Vassilevich:2004ym}. Similar difficulties appear if the metric
is matrix-valued \cite{Avramidi:2003sb}.}

The purpose of this section is to establish the small-$t$ asymptotics
of the function 
\begin{equation*}
t\to\Tr \big(L(l)R(r)e^{-tP}\big), 
\end{equation*}
where $P$ is a generalized
Laplacian, i.e. $P$ has the form
\begin{equation}
P:=-(g^{\mu\nu}\nabla_\mu\nabla_\nu+E),
\label{operD}
\end{equation}
where
\begin{align}
\nabla_\mu &:=\delta_\mu + \omega_\mu :=
 \delta_\mu + L(\lambda_\mu )-R(\rho_\mu)\,,\label{covdir}\\
E&:=L(l_1)-R(r_1)+L(l_2)  R(r_2) \,,\label{defE}
\end{align}
and $l,r,\lambda_\mu,\rho_\mu,l_i,r_i\in\Coo(\T_\Th^n)$. The 
arbitrary choice of the sign $-R$ will be justified in
($\ref{dirac}$).
One can also take more general forms of $E$ and $\omega$. For example,
$\omega_\mu$ can contain a term like $R(\rho'_\mu)L(l')$ with
some smooth $\rho'_\mu$ and $l'$. Such modifications change very
little
in our considerations below.

For that, the central asymptotic to compute is the one of the function
\begin{equation}
\label{central}
t\to\Tr\big(L(l)\,R(r)\,e^{-t\tri}\big),\quad l,r\in\Coo(\T_\Th^n).
\end{equation}
Indeed, after an expansion of the semi-group $e^{-tP}$, viewed as an
unbounded perturbation of the heat operator $e^{-t\tri}$, the only
other asymptotics we need are
\begin{equation}
\label{normal}
t\to\Tr\big(L(l)\,e^{-t\tri}\big),\quad 
t\to\Tr\big(R(r)\,e^{-t\tri}\big),
\end{equation} 
but we have shown in \cite{GI2002, Vassilevich:2002} that they 
have same asymptotics as their commutative ($\Th=0$) counterparts.

Note that the heat semi-group $e^{-t\tri}$ is trace-class, since it is
diagonal in the orthonormal basis $\{U_k\}_{k\in\Z^n}$, with eigenvalues
$e^{-t\,\vert k\vert^2}$. Of course, the same property holds for
$e^{-tP}$, as shown in the next lemma, based on a simple
application of the Duhamel expansion.

\begin{lemma}
\label{existence}
For any $r,l,\lambda_\mu,\rho_\mu,l_i,r_i\in\Coo(\T_\Th)$, 
the operator $e^{-tP}$ is trace-class for $t>0$.
\end{lemma}

\begin{proof}
We are going to use Duhamel's expansion for the semi-group generated by
$P$, viewed as an unbounded perturbation of $\tri$.
We write
$$
P=\tri-B,
$$
where $B=2g^{\mu\nu}\omega_\mu\delta_\nu+C$ with
$C=g^{\mu\nu}(\omega_\mu\omega_\nu +\omega_{\nu,\mu})+E$
and $\omega_{\nu,\mu}=L(\delta_\mu\lambda_\nu)
-R(\delta_\mu\rho_\nu)$.

{}From the Duhamel principle
$$
e^{-t(A+B)}=e^{-tA}-t\int_0^1e^{-st(A+B)}\,B\,e^{-(1-s)tA}\,ds ,
$$
we first formally write
\begin{equation}
\label{Duhamel}
e^{-t P}=\sum_{j=0}^\infty(-t)^j\,E_j(t),
\end{equation}
where
\begin{align*}
E_0(t):=e^{-t\tri} \text{ and }
E_j(t):=\int_{\tri_j} \,e^{-s_1 t\tri}\,B\,e^{-(s_2-s_1)t\tri}\cdots
B\,e^{-(1-s_j)t\tri}\,d^js.
\end{align*}
Here $\tri_j$ denotes the ordinary $j$-simplex:
\begin{align*}
\tri_j&:=\{s\in\R^j;\,0\leq s_1\leq\cdots\leq s_j\leq 1\}
\simeq \{s\in\R^{j+1};\, s_i\geq0,\,\sum_{i=0}^j s_i=1\}.
\end{align*}
We prove convergence of the expansion \eqref{Duhamel}
in the trace-norm and for reasonably small $t$: from the H\"older
inequality for Schatten classes, we have
$$
\|E_j(t)\|_1\leq\int_{\tri_j}  \|e^{-s_0 t\tri}\|_{s_0^{-1}}\,
\|B\,e^{-s_1t\tri}\|_{s_1^{-1}}\cdots
\|B\,e^{-s_jt\tri}\|_{s_j^{-1}}\,d^js.
$$
where $B=2\omega^\mu\delta_\mu+C$ and $\omega^\mu,C$ are
bounded. By functional calculus,
\begin{align*}
\|\delta_\mu\,e^{-s_it\tri}\|_{s_i^{-1}}&\leq
\|\delta_\mu\,e^{-s_it\tri/2}\|\,\|e^{- s_it\tri/2}\|_{s_i^{-1}}\\
&\leq c(g)\,(es_i t)^{-1/2}\, ( \Tr \,e^{-t\tri/2})^{s_i},
\end{align*}
using the inequality $ 
\vert \vert f(\delta_{1},\cdots,\delta_{n})\vert\vert_{op}\leq 
\vert\vert f \vert 
\vert_{\infty}$ where $f(x) = x_{\mu}e^{-x.x}$ which follows from 
$f(\delta_{1},\cdots,\delta_{n})U_{k}=f(ik)U_{k} $. So
\begin{align*}
\|E_j(t)\|_1&\leq\int_{\tri_j}  \,(\Tr \,e^{-t\tri})^{s_0}\,\Big(
\|C\|\,(\Tr \, e^{-t\tri})^{s_1}+2c(g)\sum_\mu\|\omega^\mu\|\,
(es_1t)^{-1/2}\,(\Tr \, e^{-t\tri/2})^{s_1}\Big)\cdots\\
&\hspace{3.5cm}\cdots
\Big(
\|C\|\,(\Tr \,e^{-t\tri})^{s_j}+2c(g)\sum_\mu\|\omega^\mu\|\,
(es_jt)^{-1/2}\,(\Tr \, e^{-t\tri/2})^{s_j}\Big)\,d^js\\
&\leq\Tr\big(e^{-t\tri/2}\big)\int_{\tri_j}  \,\Big(
\|C\|+2c(g)\sum_\mu\|\omega^\mu\|\,(es_1t)^{-1/2}\Big) \cdots\\
&\hspace{3.5cm}\cdots \Big(
\|C\|+2c(g)\sum_\mu\|\omega^\mu\|\,(es_jt)^{-1/2}\Big)\,d^js.
\end{align*}
Using
$$
\int_{\tri_j}\prod_{i=1}^j s_i^{-1/2}\,d^js\leq2^{j-1},
$$
the last expression can be estimated for $t\leq e^{-1}$ (since 
$s_{j}\leq 1$) by
$$
t^{-(j-1)/2}\,e^{-(j-1)/2}\,2^{j-1}
\Big(\|C\|+2c(g)\sum_\mu\|\omega_\mu\|\Big)^{j-1}
\Tr\big(e^{-t\tri/2}\big).
$$
Thus
$$
\Big\|\sum_{j=0}^\infty(-t)^j\,E_j(t)\Big\|_1\leq t\,\Tr\big(e^{-t\tri/2}\big)
\sum_{j=0}^\infty \Big(\tfrac{2}{\sqrt{e}}
(\|C\|+2c(g)\sum_\mu\|\omega_\mu\|)
\sqrt{t}\Big)^{j-1},
$$
which is finite for
$$
0<t<\frac{e}{4(\|C\|+2c(g)\sum_\mu\|\omega_\mu\|)^2}:=t_0.
$$
Finally, for $t_0 \leq t \leq 2t_0$, note that
\begin{align*}
\|e^{-tP}\|_1\leq\|e^{-(t-t_0)P}\|_1\,\|e^{-t_0P}\|,
\end{align*}
and the result follows inductively.
\end{proof}

One can probably prove this Lemma also by using some estimates involving
Sobolev spaces, cf.~\cite{Vassilevich:2002}. However, the Duhamel principle
is quite important in its own right for physical applications. We shall use
(\ref{Duhamel}) below in Sec.~\ref{sec-ncqft}. The convergence of the 
Duhamel expansion is necessary to construct the covariant perturbation
series in the approach of Barvinsky and Vilkovisky
\cite{Barvinsky:1987uw}. Note also that the Duhamel expansion has been
used to compute one-loop divergencies in a more general framework
of NCQFT, namely on Moyal plane with degenerate but non-constant $\Th$
\cite{Parthenope}.

Let us define the functional $\Sp$ on $\L(\H_\tau)$, given for a
bounded
operator $A$ by
\begin{equation}
\Sp (A):=
\lim_{t\to 0+}(4\pi t)^{n/2} \,\Tr \left( A \,e^{-t\Delta}\right).
\label{defSp}
\end{equation}
Note that this definition is not sensitive to the kernel of $\Delta$ 
(which is $\C U_{0}$), so we may change $\Delta$ in $\Delta +1$ or 
assume that $\Delta$ is invertible.

We will show that $\Sp\big(L(.)R(.)\big)$, as a functional on
$\Coo(\T_\Th^n)\x\Coo(\T_\Th^n)$, is indeed a finite and faithful 
trace in each argument, namely it vanishes whenever one of its
arguments is a commutator, see Lemma \ref{bitrace}.
This should not be surprising once one knows that 

\begin{center}
{\it $\Sp(A)$ is a
multiple of the Dixmier trace of $A(1+\tri)^{-n/2}$.}
\end{center}

\noindent Indeed, from the
knowledge of the eigenvalues of $\tri$ and the boundedness of $A$, one
has $A(1+\tri)^{-n/2}\in\L^{(1,\infty)}(\H_\tau)$, i.e., 
$\sum_{k=1}^N\mu_k\big(A(1+\tri)^{-n/2}\big)=O(\ln N)$ 
where $\mu_{k}(X)$ are the ordered singular values of $X$. It follows
then (see \cite[p.~236]{CM} with immediate modifications) that
$$
\Tr_\omega\big(A(1+\tri)^{-n/2}\big)= \frac{1}{\Ga(n/2+1)}\,
\lim_{t\to0} \,t^{n/2} \,\Tr 
(A\,e^{-t\tri})= \frac{1}{(4\pi)^{n/2}\Ga(n/2+1)}\,\Sp(A).
$$
However, we leave this feature now since our goal is to find an
algorithm to obtain analytic expressions for the heat
coefficients. For that, Dixmier-trace technology is not so helpful.

{}From relations \eqref{rel} and using the orthonormal basis
$\{U_k\}_{k\in\Z^n}$ to compute the trace, we find for 
$l=(2\pi)^{-n/2}\sum_{q_{1}}l_{q_{1}}U_{q_{1}}$, 
$r=(2\pi)^{-n/2}\sum_{q_{2}}r_{q_{2}}U_{q_{2}}$ 
in $\Coo(\T^n_\Th)$,

\begin{align*}
\Tr\big(L(l)\,R(r)\,e^{-t\tri}\big)
&=\sum_{k \in\Z^n}\tau\big(U_k^*\,L(l)\,R(r)\,e^{-t\tri}\,U_k\big)
=\sum_{k \in\Z^n}e^{-t\vert k\vert^2}\tau\big(U_k^*\,l\,U_k\,r\big) \\
&=(2\pi)^{-n}\sum_{k,\,q_1,\,q_2\in\Z^n}
e^{-t\vert k\vert^2}\,l_{q_1}\,r_{q_2}\,\tau\big(
U_{-k}\,U_{q_1}\,U_k\,U_{q_2}\big) \\
&=(2\pi)^{-n}\sum_{q,\,k\in\Z^n}l_q\,r_{-q}\,
e^{-ik.\Th q}\,e^{-t\vert k\vert^2},
\end{align*}
which after Poisson resummation reads
\begin{align}
\label{1847}
\Tr\big(L(l)\,R(r)\,e^{-t\tri}\big)&=\sqrt g \,(4\pi t)^{-n/2}
\sum_{q,\,k\in\Z^n}l_q\,r_{-q}\,e^{-|\Th q-2\pi k|^2/4t}\nonumber\\
&=\sqrt g\sum_{q,\,k\in\Z^n}l_q\,r_{-q}\,K(t,\Th q-2\pi k)
\end{align}
where $K(t,x):=(4\pi t)^{-n/2} \, e^{-|x|^2/4t}$ is the heat kernel
of $\R^n$ with metric $g^{\mu\nu}$.

To proceed further, we need to impose some restrictions
on the matrix $\Theta$. 
Whereas it does not for the asymptotics of \eqref{normal},
the number-theoretical aspect of
$\Th$ has huge consequences for the asymptotics of \eqref{central}.

\noindent To explain what this is about, let us review 
what happens in the (nondegenerate) two-dimensio-nal case, where
$\Th=\th\begin{pmatrix}0&-1\\1&0\end{pmatrix}$, with $\th$ an
arbitrary real number. (Actually, up to an isomorphism, 
the range of $\th$ can
be reduced to the interval $[0,\thalf]$.)
In this case, there are two distinct situations to consider
\cite{UV/IR}. When $\th$ is rational (relatively to $2\pi$), then
in the sum \eqref{1847}, only terms with $q$ multiple of
$\th$-denominator, 
will contribute to the small-$t$
asymptotics. When $\th$ is irrational (again relatively to $2\pi$), one
can guess that only the zero-mode will contribute to the small-$t$
asymptotics of $\Sp\big(L(l)R(r)\big)$. This is indeed true, provided
one has a control on the sum
\begin{equation}
\label{2D}
\sum_{k\in\Z^2}\sum_{0\ne q\in\Z^2}l_q\,r_{-q}\,\frac{e^{-|\th q-2\pi
    k|^2/4t}}{(4\pi t)^{n/2}}\,,
\end{equation}
i.e., provided one can measure how far from rationals $\th$ is,
since $|\th q-2\pi k|$ could be in principle arbitrarily close to
$0$.\\
This control is precisely given by a Diophantine condition.
\begin{definition}
A number $\th$ is said to satisfy a Diophantine condition
(relatively to $2\pi$) if
there exists two constants $C>0$, $\beta\geq0$ such that for all 
$q\in\Z^*$
\begin{align}
 \|\th q\|_\T:=\inf_{k\in\Z}|\th q-2\pi k|\geq
  \frac{C}{|q|^{1+\beta}} \Longleftrightarrow \,
  \vert 1 -\cos (\th q)\vert \geq 
  \frac{2C^2}{\vert q \vert ^{1+2\beta+\beta^2}} \label{Diophcond}
\end{align}
\end{definition}
In fact, $\inf_{p\in\Z} 
\vert \th q-p \vert \leq \vert \sin(\pi \th q) \vert:$ actually, 
for a given $q$, their exists an integer $p_{0}$ such that 
$ \vert \th q -p_{0}\vert \leq\tfrac 12$ 
and since we work under modulus, we may assume that $0\leq  \th q-p_0 \leq 
\tfrac 12$. Since $(\sin x)/x$ is decreasing on $[0,\tfrac\pi 2]$, 
we get $\sin(\pi(\th q-p_{0}))\geq 2(\th q-p_{0})$ and the above 
equivalence by taking the square of the inequality with 
$\sin^2(\tfrac{\th q}{2})=\tfrac 12 (1-\cos(\th q))$.

In other words, this condition states that the inverse torus norm of
$\th q$ is a temperate distribution over $\Z^*$. This is
exactly what we need since in equation \eqref{2D}, the complex
coefficients $l_q$, $r_q$ are of Schwartz-class by the smoothness
assumption. Note that this condition is not too restrictive since the
set of irrational numbers satisfying a Diophantine condition is of
full Lebesgue measure.
 
In the general Poisson case, 
one can always change the coordinates on $\T^n$, $y=Bx$ with a constant
matrix $B$, so that the new coordinates $y$ are $2\pi$-periodic again,
and the Poisson matrix $\bar \Th = B^T \Th B$ becomes   
\begin{equation}
\label{bel}
\bar \Th=0_{l} \, \bigoplus_{\{i\,:\,\th_{i}\in 2\pi\mathbb{Q}^*\}}\th_{i}
\begin{pmatrix}0&1\\-1&0\end{pmatrix} \, \bigoplus_{\{j\,:\,\th_{j}\in 
    \R\backslash 2\pi\mathbb{Q}\}}\th_{j}
\begin{pmatrix}0&1\\-1&0\end{pmatrix} 
\end{equation}
where $0_{l}$ is the zero matrix of size $l$ and 
$n=l+m_1+m_2$. Here $m_1$  is the size of the
rational part (with $\th_i=p_i/q_i$, $i=1,\cdots,m_1$) 
and $m_2$ of the irrational one. We define
then $\mathcal{Z}=\Z^l\x q_1\Z\x\cdots
\x q_{m_1}\Z\x\{(0,\cdots,0)\in\Z^{m_2}\}$. The rest of $\mathbb{Z}^n$
is denoted by $\mathcal{K}$ and is splited into two parts,
$\mathcal{K}^{\rm
  per}\simeq \{(k_1,\cdots,k_{m_1})\in\N^{m_{1}},\,1\leq k_i\leq q_i-1,\,
i=1,\cdots,m_1\}$ and $\mathcal{K}^{\rm inf}\simeq\Z^{m_2}\setminus\{0\}$.
$\mathcal{K}^{\rm per}$ is a finite set which lays in an
$\mathbb{R}$-linear space generated by $\mathcal{Z}$. We shall omit
the bar over $\Th$ in what follows. 

One can avoid the use of a particular coordinate system. 
Then $\mathcal{Z}$ is defined as the set of $q\in \mathbb{Z}^n$ such that 
$(2\pi)^{-1} \Theta q \in \mathbb{Z}^n$. This definition reveals
the actual meaning of this set. One can also figure out how to give
coordinate independent definitions of $\mathcal{K}^{\rm per}$ and
$\mathcal{K}^{\rm inf}$.

\quad

Now we are ready to formulate our restriction on $\Theta$.
We assume the following:

{\it The matrix $\Theta$ satisfies a Diophantine
condition
with respect to $\mathcal{K}^{\rm inf}$, i.e., there are two
positive constants $C$ and $\beta$ such that
\begin{equation}
\inf_{k\in \mathbb{Z}^n} | \Theta q - 2\pi k| \ge \frac {C}{ |q|^{1+\beta}}
\quad \mbox{for all }\ q\in \mathcal{K}^{\rm inf}.
\label{Diop}
\end{equation}}
In the standard definition of a Diophantine condition for an $l$-tuple
of real numbers, one assumes that $1+\beta \geq l$. Our consideration is
valid in a more general case for any positive $1+\beta$, so we do not need
this restriction as far as such $\Th$ exist; note, for instance, that there
exists a set of full Lebesgue measure of $l$-tuples of Roth type, that is of
$\alpha \in \R^l$ such that for all $\epsilon>0$, there exists $C_\epsilon$
with $\inf_{k \in \Z^l} | \alpha q - 2\pi k| 
\ge \tfrac{C_\epsilon}{ |q|^{n+\epsilon}}$, see \cite{Herman}.

We remark that it does not matter whether one uses the metric $g^{\mu\nu}$ 
or the normalized diagonal metric in norm-values of (\ref{Diop}) 
since it can be absorbed in the constant $C$.

With this restriction on $\Theta$ we can prove the following formula
which governs the asymptotic behavior of the trace at $t\to 0^+$. 
\begin{equation}
 \Tr \left( L(l) R(r) e^{-t\Delta}\right) = \frac{\sqrt g}{(4\pi
t)^{n/2}}
\sum_{q\in \mathcal{Z}}
l_q \, r_{-q} + \mbox{ e.s.t.}
\label{asym1}
\end{equation}
where e.s.t. denotes some exponentially small terms in $t$, i.e., the
terms
which vanish faster than any power of $t$ as $t\to 0^+$.
The proof, which is somewhat technical, is postponed to Appendix~\ref{appA}.

Equation (\ref{asym1}) immediately yields the following Theorem.

\begin{theorem} 
Assume $\Theta$ satisfies condition (\ref{Diop}).
Then for any $l,r\in\Coo(\T^n_\Th)$,
\begin{equation}
\Sp \big(L(l) R(r)\big)= \sqrt g \sum_{q\in \mathcal{Z}}
l_q \, r_{-q}. \label{eSp}
\end{equation}
\end{theorem}

The explicit expression (\ref{eSp}) makes it possible to show rather
directly that
(\ref{defSp}) indeed defines a trace in each variable: it vanishes 
whenever $l$ or $r$ is a commutator:

\begin{corollary} Let $l, r, s \in \Coo(\T^n_\Th)$. Then,
\begin{equation*}
\label{bitrace}
\Sp \big(L(l) R([r,s])\big)=0.
\end{equation*}
\end{corollary}
\noindent
\begin{proof}
{}From commutation relations \eqref{rel1}, we find
$$
[r,s]=-2i (2\pi)^{-n}\sum_{k,\,q\in\Z^n}r_k\,s_q\,
\sin(\thalf k.\Th q)\,U_{k+q}\,,
$$
and thus
$$
[r,s]_k=2i(2\pi)^{-n/2}\sum_{q\in\Z^n}r_{k-q}\,s_q\,\sin(\thalf q.\Th
k),
$$
which is zero whenever $k\in\mathcal{Z}$ since it is equivalent to
$(2 \pi)^{-1}\Th k\in\Z^n$. 
\end{proof}

\begin{remark}
We have the following relations between the functional $\Sp$
and the trace $\tau$:
$$
\Sp\big(L(a)\big)=\Sp\big(R(a)\big)=({\rm vol}\, \T^n)\,\tau(a),
$$ 
for all $a\in\Coo(\T^n_\Th)$ and any $\Th$, and when
$\mathcal{Z}=\{0\}$ (i.e. pure Diophantine case), then
$$
\Sp\big(L(l)\,R(r)\big)=({\rm vol}\, \T^n)\, \tau(l)\,\tau(r),
$$
which makes transparent the statement of the corollary.
\end{remark}

This completes our study of the trace (\ref{defSp}). Below we present
several relations similar to (\ref{asym1}) which will be used in the
next section. One can show (see Appendix \ref{appA}) that
\begin{equation}
 \Tr \left( [L(l) R(r)]^\mu \delta_\mu e^{-t\Delta}\right) = 0 + 
\mbox{ e.s.t.},
\label{asym2}
\end{equation}
where the notation $[L(l) R(r)]^{\mu_1 \dots \mu_m}$ means that the
vector indices are distributed between $l$ and $r$. 
For higher derivatives we have
\begin{align}
\Tr \left( [L(l) R(r)]^{\mu_1 \dots \mu_m} \,
\delta_{\mu_1} \dots \delta_{\mu_m} \,e^{-t\Delta}\right) 
&= \sum_{k\in\Z^n}\tau\big(U_k^*\,[L(l) R(r)]^{\mu_1 \dots \mu_m} \,
\delta_{\mu_1} \dots \delta_{\mu_m}\,
e^{-t\Delta}\,U_k\big)\nonumber\\
&= i^m\, \sum_{k\in\Z^n}k_{\mu_1} \dots k_{\mu_m} 
\tau\big(U_k^*\,[L(l) R(r)]^{\mu_1 \dots \mu_m} \,
e^{-t\Delta}\,U_k\big)\nonumber\\
&=: i^m \, G^{(m)}_{\mu_1 \dots \mu_m} 
\Tr  \left( [L(l) R(r)]^{\mu_1 \dots \mu_m} 
\,e^{-t\Delta}\right) ,\label{moreder}
\end{align}
One can calculate the tensors $G^{(m)}$
by varying (\ref{asym1}) or (\ref{asym2}) with respect to the
metric $g^{\mu\nu}$ (this is a standard way to include derivatives
in the heat trace expansion, cf.~\cite{Branson:1997ze}.)
All $G^{(2j+1)}$ are 
exponentially small and can be neglected in our analysis. For even
$m$, corresponding tensors $G^{(m)}$ are obtained from the following
recursion relation:
\begin{equation}
G^{(2p+2)}_{\mu\nu\mu_1\dots\mu_{2p}}
= -\frac 1t \, \frac{\delta}{\delta g^{\mu\nu}}
G^{(2p)}_{\mu_1\dots\mu_{2p}}, \label{recur}
\end{equation}
with $G^{(0)}=\sqrt{g}$. One has to take into account that
$g^{\mu\nu}$
is symmetric, so that not all of the components are indeed
independent,
and
\begin{equation}
 \frac{\delta}{\delta g^{\mu\nu}} g_{\rho\sigma} =
-\tfrac 12 (g_{\mu\rho}g_{\nu\sigma} + g_{\mu\sigma}g_{\nu\rho}).
\label{varg}
\end{equation}
For example,
\begin{align}
G^{(2)}_{\mu\nu}&=\tfrac {\sqrt g}{2t}  \, g_{\mu\nu} \,,\nonumber\\
G^{(4)}_{\mu\nu\rho\sigma}&=\tfrac {\sqrt g}{4t^2} \,
(g_{\mu\nu}g_{\rho\sigma} + g_{\mu\rho}g_{\nu\sigma} +
g_{\mu\sigma}g_{\nu\rho}). \label{G24}
\end{align}

\subsection{Heat trace asymptotics for generalized Laplacians}

{}From expression \eqref{operD} of generalized Laplacian 
$P=-g^{\mu\nu}\nabla_\mu\nabla_\nu-E$, we
need also to define the associated curvature:

\begin{equation*}
\widehat \Omega_{\mu\nu} := \nabla_\mu  \nabla_\nu - \nabla_\nu
\nabla_\mu
=L(\Omega_{\mu\nu}^L)-R(\Omega_{\mu\nu}^R),
\end{equation*}
and higher ``covariant derivatives'' of $E$ and $\widehat \Omega$ as 
repeated commutators with $\nabla$. For example, $E_{;\mu} :=
[\nabla_\mu,E]$, $E_{;\mu\nu}:=[\nabla_\nu, E_{;\mu}]$.

Actually, there are two covariant derivatives, 
$\nabla^L_\mu :=\delta_\mu + L(\lambda_\mu)$
and $\nabla^R_\mu :=\delta_\mu - R(\rho_\mu)$
and two gauge symmetries in the problem. One of these symmetries
acts on ``left'' fields $\lambda_\mu$, $l_{1},l_{2}$, the  
other acts on ``right'' fields $\rho_\mu$, $r_{1},r_{2}$.
The gauge group has a direct product structure.
Explicit expressions for the symmetry transformations can be 
found in \cite{Vassilevich:2005vk}. The functional 
$\Sp$ of any polynomial of $E$, $\widehat \Omega$ and derivatives is
gauge invariant. Full invariance will mean that also 
all vector indices are contracted in pairs. This is precisely the
class
of invariants which will appear in the heat trace asymptotics.
Due to the product structure of the gauge group there are gauge
invariants which do not belong to the class we have just described.
For example, $\Sp \big(L(l_1)\big)$ is such an invariant.

We will see that
in spectral action picture, there is only one gauge group, namely
the automorphism group of the algebra but lifted to the spinor bundle
via the charge conjugation operator. In such application, the
representation really looks like the adjoint one (see
Sec.~\ref{spectralaction}).

We shall need the notion of {\it canonical mass dimension}.
We assign canonical mass dimension $2$ to $E$ and $\widehat \Omega$,
and canonical mass dimension $1$ to each derivative.
Canonical mass dimension of any monomial is the sum of canonical
mass dimensions of all factors.

The heat trace asymptotic is then given by the following

\begin{theorem}
\label{theorem1}
Let $P$ be as defined above. Then

i)
There is a full asymptotic expansion of the heat
trace
\begin{equation}
\Tr \left( L(l) R(r) e^{-tP}\right)\, \underset{t\to 0^+}{\sim} \, 
\sum_{k=0}^\infty
\,a_k(l,r;P) \, t^{(k-n)/2}\, . \label{asymptotex}
\end{equation}

ii)
The coefficients $a_k$ can be expressed as 
\begin{equation}
\label{aoriginal}
a_k(l,r;P) = \sum_\alpha \, b_\alpha \, \Sp
\big( L(l) R(r) \mathcal{A}_\alpha \big) ,
\end{equation}
where $\mathcal{A}_\alpha$ are independent invariant 
free polynomials of canonical mass dimension $k$ of
$E$, $\widehat \Omega$ and their covariant derivatives.
The numbers $b_\alpha$ are constants.
The odd-numbered coefficients $a_{2j+1}$ vanish. 

iii)
Values of $b_\alpha$ are uniquely defined by considering
``pure left'' ($\rho_\mu =r_1=r_2=0$, $r=1$) or ``pure right'' 
($\lambda_\mu =l_1=l_2=0$, $l=1$) cases. In particular,
\begin{eqnarray}
&&a_0(l,r,P)= (4 \pi)^{-n/2}\, \Sp \big( L(l)R(r)\big) ,\label{a0lr}\\
&&a_2(l,r,P)= (4 \pi)^{-n/2}\, \Sp \big( L(l)R(r) E \big)
,\label{a2lr}\\
&&a_4(l,r,P)= (4 \pi)^{-n/2}\, \tfrac{1}{12} \,\Sp \big( L(l)R(r)
(6 E^2 +2{E_{;\mu}}^\mu+ \widehat
\Omega^{\mu\nu}\widehat\Omega_{\mu\nu} )\big). 
\label{a4lr}
\end{eqnarray}
\end{theorem}

\begin{proof}
Existence of the trace follows from Lemma \ref{existence}. The proof
of
the second statement (below) can be considered as a constructive proof
for the existence of the asymptotic expansion.

To evaluate the asymptotic behavior of the trace on the left hand side
of (\ref{asymptotex}) we use the canonical basis $\{U_k\}$ of
$\H_\tau$. 
This is a standard procedure used in quantum field theory for a long
time (cf.~\cite{Nepomechie:1984wt}), which was recently applied to
noncommutative
theories \cite{Vassilevich:2002,Vassilevich:2005vk}: one first
factors out a global $e^{-t|k|^2}$ term,
\begin{align}
\Tr \left( L(l) R(r) e^{-tP}\right) &=
\sum_k\tau\Big(U_k^*\, L(l)\, R(r)\, e^{-tP}\,U_k\Big) \nonumber \\
&=\sum_k e^{-t\vert k\vert^2}\tau\Big(U_k^*\, L(l)\, R(r)\,
e^{t((\nabla^\mu-ik^\mu)(\nabla_\mu-ik_\mu)+2ik^\mu(\nabla_\mu-ik_\mu)+E)}
\,U_k\Big). \label{toex}
\end{align}
Then, one expands the exponential in (\ref{toex}) as a
power series in $E$ and $(\nabla -ik)$. As a result, one gets a sum 
of monomials of the form 
\begin{equation}
\sum_k e^{-t\vert k\vert^2}k_{\mu_1}\cdots
k_{\mu_m}\,\tau\big(U_k^*\,L(l)\,R(r)\,F(E,(\nabla-ik))^{\mu_1\cdots\mu_m}\,
U_k\big).
\label{1640}
\end{equation}
We stress that it is important that each $\nabla$ appear
in the combination $(\nabla -ik)$.
We take in $F$ all $(\nabla -ik)$ one by one starting with the
rightmost
$(\nabla -ik)$ and push them to the right. Being commuted through an
operator of the type $L(f)R(h)$, $(\nabla -ik)$ replaces the
functions 
$f,h$ in this operator by their derivatives, e.g., $(\nabla -ik) L(f) =
L(f) (\nabla -ik) + L(\nabla^L f)$, where only the Leibniz rule 
satisfied by the derivations $\delta_\mu$ has been used. 
When $(\nabla -ik)$ hits $U_k$,
it becomes an operator of (left and right) multiplication by the
connection,
i.e., $(\nabla_\mu -ik_\mu) \,U_k=
\big(L(\lambda_\mu)-R(\rho_\mu)\big) \,U_k$.
In this way, one can remove all derivative operators (which are
replaced
by left and right multiplication operators) 
and all momenta $k$ from the expression inside the NC-torus trace 
in (\ref{1640}). Therefore, one can apply (\ref{moreder}) to obtain 
instead of (\ref{1640}) the following expression
\begin{equation}
G^{(m)}_{\mu_1 \dots \mu_m} 
\sum_k e^{-t\vert k\vert^2}
\tau
\big(U_k^*\,L(l)\,R(r)\,F(E,(\nabla-ik))^{\mu_1\cdots\mu_m}\,U_k\big).
\label{1702}
\end{equation}

Let us now evaluate the power of $t$ corresponding to each 
monomial. If $F$ contains an $N_E$-th power of $E$, an $N_P$-th power 
of $(\nabla -ik)^2$ and an $N_K$-th power of $2ik^\mu (\nabla_\mu
-ik_\mu)$,
then $F$ itself contains $t^{N_E+N_P+N_K}$. Explicit multipliers
$k_\mu$ which we put in front of $F$ in (\ref{1640}) come from
$2ik^\mu (\nabla_\mu -ik_\mu)$ only. Consequently, $m=N_K$. For 
odd $m$, the tensors $G^{(m)}$ vanish up to exponentially small terms,
while for an even $m$ the tensor $G^{(m)}$ is proportional to
$t^{-m/2}$. The sum over $k$ brings another $t^{-n/2}$. Altogether,
we have $t^{N_E + N_P + N_K/2 - n/2}$. This means that such monomials
contribute to the coefficient $a_p(l,r,P)$ with $p=2N_E + 2N_P + N_K$,
which are precisely the canonical mass dimensions of the monomials as
defined above. Besides, $N_K$ should be even. Consequently, odd
numbered heat kernel coefficients $a_{2j+1}$ vanish.

\quad

Now we have to prove that the heat kernel coefficients are of the form
declared in the Theorem, i.e., that they are invariant polynomials
constructed from $E$, $\widehat \Omega_{\mu\nu}$ and their
derivatives.
We have already proved that the expression inside the trace $\tau$ 
in (\ref{1702}) is in fact a multiplication operator (i.e., a
combination of left and right regular representation operators) 
which does not contain $k$. Together with gauge invariance of the 
heat trace this could have been enough to get the statement. 
However, since the gauge group has a
product structure, there are more gauge invariants than we expect to
find in
the heat trace asymptotics. Let us collect all
monomials $F_a (E,(\nabla -ik))^{\mu_1 \dots \mu_m}$ of a given (even)
canonical mass dimension $p$ which appear in the expansion of the
exponential of (\ref{toex}), and consider the sum
\begin{equation}
\tau \biggl( U^*_k\, L(l)\, R(r)\,
\sum_a G^{(m)}_{\mu_1 \dots \mu_m}  F_a (E,(\nabla -ik))^{\mu_1 \dots
\mu_m} U_k \biggr).
\label{1612}
\end{equation}
As we have demonstrated above, $F_a$ are free polynomials of $E$,
$\omega$ and their derivatives $\delta_\mu E$, $\delta_\mu \omega_\nu$
etc. The same procedure as above can be carried out for an arbitrary
Laplace type operator $\bar P$ acting on smooth sections of an arbitrary
(non-abelian) vector bundle over the {\it commutative} torus $\T^n$. 
The free polynomials of the endomorphism $\bar E$, the connection
$\bar \omega_\mu$, which characterize $\bar P$, and their derivatives
are in one-to-one correspondence with the polynomials in (\ref{1612}).
In the case of $\bar P$ we know that all terms can be recombined
into covariant derivatives and field strengths thus giving
standard heat kernel coefficients.
This is a purely combinatorial statement, which does not depend on
the nature of $\nabla$ (or $\bar \nabla$) and 
$E$ (or $\bar E$). Therefore, the same recombination can be done
also in the noncommutative case considered here. This completes the
proof of the second assertion.

The third point is easy, it simply means that independent invariants remain
independent when reduced to ``pure left'' or ``pure right'' cases. The
coefficients in front of these invariants can therefore be read off from 
``pure left'' heat kernel coefficients \cite{Vassilevich:2002}
on the torus. This includes (\ref{a0lr}), (\ref{a2lr}), (\ref{a4lr})
and even $a_6$ which is not given explicitly in 
the present work.
\end{proof}

The interested reader can calculate also higher terms in the heat
trace asymptotics by using the expressions for $a_8$ 
\cite{Avramidi:1990je} and $a_{10}$ \cite{vandeVen:1997pf} 
obtained in the commutative case.

\section{Toward the asymptotics for toric 
noncommutative manifolds}

This section is devoted to the study of the
asymptotic \eqref{asym1} in a more general setting of noncommutative
spaces. We will concentrate here on toric noncommutative manifolds,
$\Coo(M_\Th)$, also called periodic isospectral deformations (the
aperiodic case \cite{GIV}, akin to Moyal plane, will be studied
elsewhere).
This class of quantum spaces can be thought as a curved space
generalization of NC-tori. They were originally defined by Connes and
Landi \cite{ConnesLa} (from cohomological considerations) 
within a twisted product approach (that we will 
follow here) and later by Connes and Dubois-Violette 
\cite{ConnesDV} in a more intrinsic way via fixed-point algebra
techniques.

We first recall the definition of $\Coo_c(M_\Th)$: 
let $(M,g)$ be a Riemannian (compact or not) $n$-dimensional manifold
without
boundary. Consider $\a:\T^l\to\Isom(M,g)$, a smooth isometric action
of a
$l$-torus on $M$, typically given by the maximal abelian subgroup of
the isometry group of the manifold (the interesting class is $l\geq
2$).
This action induces a spectral (Peter--Weyl) decomposition of 
any smooth function with compact support $f\in\Coo_c(M)$
$$
f=\sum_{r\in\Z^l}f_r, \sepword{such that} \a_z(f_r)=e^{-ir.z}f_r,
\,\forall z \in\T^l,
$$
where the action by automorphism of the $l$-torus on $\Coo_c(M)$ 
(also denoted $\a$) is given by $(\a_z f)(p):=f(\a_{-z}(p))$.

It is important to notice that this
expansion is convergent in the sup-norm $\|.\|_\infty$ (in fact
$\|f_r\|_\infty$ is a Schwartz sequence for $f\in\Coo_c(M)$.)

By analogy with the noncommutative torus, given a skewsymmetric $l\x
l$ matrix $\Th$, one can deform the algebra
$\Coo_c(M)$ to a noncommutative one $\Coo_c(M_\Th)$, defining the
following twisted product on pairs of homogeneous elements 
\begin{equation}
\label{StarP}
f_r\Mop g_s=e^{-\frac{i}{2}r.\Th s}\,f_r.g_s.
\end{equation}
It should be clear that this product can also be realized via the
Rieffel star-product \eqref{Rieffel} associated to the action $\a$ 
\cite{ReiffelDefQ}. In this setting, the natural trace $\tau$ of this
algebra is the integral with Riemannian volume form $\mu_g$
\begin{equation}
\label{trrace}
\tau(.)=\int_M (.)\,\mu_g ,
\end{equation}
and all first-order differential
operators which commute with the action $\a$ form a Lie algebra of
derivations. 
 
On the Hilbert space $\H$ of square integrable function on $M$ with
Riemannian volume form $\mu_{g}$, one 
defines left and right twisted multiplication operators $L(l)$,
$R(r)$, $l,r\in\Coo_c(M)$, by
$$
L(l)\psi=l\Mop\psi, \quad R(r)\psi=\psi\Mop r, \sepword{for all}
\psi\in\H.
$$
Because the Peter--Weyl expansion is sup-norm convergent, those
operators
are bounded for (at least) smooth compactly supported functions.   

For any isometric $\R^l$-action on a Riemannian manifold, 
from the expression of the kernel of the operators $L(l)$,
$R(r)$ (see for instance \cite{UV/IR}) and using the heat kernel 
$K_t(p,p')$ of the scalar Laplacian associated with the metric $g$
and its 
volume form $\mu_{g}$, one can
compute
$$
\Tr\big(L(l)\,R(r)\,e^{-t\tri}\big)=(2\pi)^{-l}\int_M\mu_g(p)
\int d^ly \,d^lz\,l(p)\,r(\a_z(p))\,K_t\big(\a_{-\Th y}(p),p\big).
$$
After a Peter--Weyl expansion of the functions $l,r$, this reads

\begin{align}
\label{genetrace}
\Tr\big(L(l)\,R(r)\,e^{-t\tri}\big)&=\sum_{q\in\Z^l}\int_M\mu_g(p)\,
l_q(p)\,r_{-q}(p)\,K_t\big(\a_{\Th q}(p),p\big).
\end{align}

    For a large class of manifolds (see \cite{GIV,DaviesHeatKer} 
for a review of sufficient
conditions), the behavior of the off-diagonal heat kernel is
controlled by the geodesic distance function:
\begin{equation}
\label{eq:HK}
\frac{1}{(4\pi t)^{n/2}}\,e^{-d_g^2(p,p')/4t}\leq K_t(p,p')\leq C
\frac{1}{(4\pi t)^{n/2}}\,e^{-d_g^2(p,p')/4(1+c)t},
\end{equation}
where $d_g$ is the geodesic distance and $C,c$ are positive constants.

This estimate and the fact that the metric on the orbits of the torus
action is constant (since $\T^l$-action is isometric) 
show that the previous discussion on the role of
the arithmetic nature of the deformation parameters applies also in
this setting (since the geodesic distance on the
orbits is the torus one). But in this framework this is not the
end of the story.

Indeed, such an action is not necessarily free (for example it is for
NC-torus but not for Connes--Landi spheres), that is, there may
exist fixed or rather singular points for the action. For instance,
on a neighborhood of a fixed point,
we see that in the integral \eqref{genetrace}, we are left
with the heat kernel on the diagonal (i.e., the dumping factor of the
exponential of geodesic distance disappears). This has certainly 
some consequences for the power-$t$ expansion. In view of
\eqref{eq:HK}, note that the lack of
freedom for the action can be rephrased in term of non-local
integrability of the function $p\mapsto d_g^{-2}(\a_y(p),p)$ for
certain $0\ne y\in\T^l$, in the neighborhood of singular points.
 
At this level of generality, we are only able to treat the free
torus-action case, where one can easily derive the asymptotic of
\eqref{genetrace}. From previous techniques, the estimate \eqref{eq:HK}
and under the Diophantine assumption \eqref{Diop}, one gets 
\begin{align}
 \Tr \left( L(l) R(r) e^{-t\Delta}\right) &= \frac{1}{(4\pi t)^{n/2}}
\sum_{q\in \mathcal{Z}}\int_M\mu_g(p)\,
l_q(p) \, r_{-q}(p)\,k_t(p,p) + \mbox{ e.s.t.}  \nonumber\\
&= \frac{1}{(4\pi t)^{n/2}}\sum_{k=0}^\infty\,t^k\,\sum_{q\in \mathcal{Z}}
\int_M\mu_g(p)\,l_q(p) \, r_{-q}(p) \, a'_{2k}(p) +\mbox{ e.s.t.},
\label{asym33}
\end{align}
where $a'_{2k}(p)$ are the local heat kernel coefficients for the
scalar Riemannian Laplacian.

For a non-free torus action, it seems to be difficult to outstrip the
qualitative level in general, i.e., to express the asymptotic of 
\eqref{genetrace}
in terms of geometric invariants. We will instead treat the (quite
simple but non-trivial) example of the ambient space of Connes--Landi
3-sphere $\Sf^3_\th$ \cite{ConnesLa}. 

One standard way to construct this ambient space goes as follow. 
One parameterizes $\R^4$ (with
standard metric) in spherical $(\phi_1,\phi_2,\psi)$, $\phi_i\in\T$,
$\psi\in[0,\pi/2]$ (with non-trivial boundary conditions) and radial
$R\in[0,+\infty[$ coordinates. That is to say, in terms of Cartesian
coordinates:
\begin{align*}
&x_1=R\,\cos\psi\,\cos\phi_1,\quad x_2=R\,\cos\psi\,\sin\phi_1,\\
&x_3=R\,\sin\psi\,\cos\phi_2,\quad x_4=R\,\sin\psi\,\sin\phi_2.
\end{align*}
Then, one twists the product via the $\T^2$-action
$$
y.(R,\psi,\phi_1,\phi_2)=(R,\psi,\phi_1+y_1\Mod2\pi,\phi_2+y_2\Mod2\pi),
\quad y\in\R^2.
$$
In other words, we are mapping the commutative generators
$u_i=e^{2i\pi\phi_i}$, $i=1,2$, to those of the noncommutative
2-torus (of course $\Sf^3_\th$ is obtained by imposing the sphere
relation on the generators).

For the question of the asymptotic of \eqref{genetrace}, it is more
convenient to move to another coordinate system. It allows to
identify the ambient space of $\Sf^3_\th$ with the ambient space of
$\T^2_\th$. \\
This is achieved by setting
$$
\rad_1=R\,\cos\psi,\quad \rad_2=R\,\sin\psi,
$$
which leads to a parameterization of $\R^4$ in double polar coordinates
$(\rad_1,\phi_1;\rad_2,\phi_2)$. 
Thus, it corresponds to twist the product of the commutative 
algebra $\SS(\R^4)$ via
the action of $\T^2$ given by the two $SO(2)$-rotations 
(which generate the maximal compact Abelian subgroup of the 
isometry group of $\R^4$). In such a case, 
the only interesting situation is when 
$\Th=\th\begin{pmatrix} 0&1\\-1&0\end{pmatrix}$, with $\th$ irrational.

We make the Fourier transform in the two angular directions and leave
two radial coordinates $\rad_{1},\rad_{2}$ as they are. From the
expression of the heat kernel
of $\R^4$ parameterized by ($\rad_{1}\cos\phi_{1},\rad_{1}\sin
\phi_{1}, 
\rad_{2}\cos\phi_{2},\rad_{2}\sin \phi_{2}$) and action 
$(\phi'_{1},\phi'_{2}).(\rad,\phi_{1},\phi_{2}):=
(\rad,\phi_{1}+\phi'_{1} \bmod 2\pi,\phi_{2}+\phi'_{2} \bmod 2\pi)$,
we have
$$
K_{t}\big(\a_{-\Th q}(p),p\big)=\frac{1}{(4\pi t)^2} \,
e^{-\big(\rad_{1}^2(1-\cos \th q_{2})+\rad_{2}^2(1-\cos\th 
q_{1})\big)/2t},
$$
and we obtain for
\eqref{genetrace} 
\begin{equation}
\Tr \big(L(l) R(r) e^{-t\Delta}\big)= 
\frac c{t^2} \int_{\R^+\times \R^+} d^2\rad\, \rad_1 \rad_2  
\sum_{q\in \mathbb{Z}^2}
l_{q}(\rad)\,r_{-q}(\rad)\, 
e^{
-\big(\rad_1^2 (1-\cos \theta q_2) + \rad_2^2 (1-\cos \theta 
q_1)\big) /2t}.
\label{1856}
\end{equation}

The term with $q=0$ in (\ref{1856}) gives our ``standard'' result.
Other terms give (in the asymptotics) contributions from the
``singularities'' in $\rad_1=0$, $\rad_2=0$.

There is a relation between oscillations of a smooth function
in the angular directions and its behavior near the origin of the
coordinate system $\rad =0$. Consider a smooth complex function $\psi$
on $\mathbb{R}^2$. Let us restrict it to the unit disc in
$\mathbb{R}^2$.
We are going to expand $\psi$ is a series of eigenfunctions of
the Laplace operator on the disc. Consider a polar coordinate system
$(\rad,\phi)$ centered at the origin. 
Let $\psi^0$ be a restriction
of $\psi$ to the boundary of the disc, and 
$\psi^0(\phi)=\sum_l \psi^0_l e^{il\phi}$. Then $\psi^0_\rad (\phi):=
\sum_l \psi^0_l e^{il\phi}\rad^{\vert l \vert}$ is a (smooth)
zero mode of the Laplacian, and 
$\tilde\psi (\rad,\phi):=\psi(\rad,\phi)-\psi^0_\rad(\phi) $
is a smooth function satisfying Dirichlet boundary conditions on
the boundary of the disc. $\tilde \psi$ can be expanded in a sum
of non-zero eigenfunctions of the Laplacian, which are
$e^{il\phi} J_{\vert l \vert} (\rad \lambda)$, where $J_{\vert l
\vert}$
is the Bessel function, and the
eigenvalues $\lambda$ are
defined by the boundary condition $J_{\vert l \vert} (\lambda)=0$.
The Taylor expansion for $J_{\vert l \vert}$ around $\rad =0$ starts
with $\rad^{\vert l \vert}$ and contains the powers
$\rad^{\vert l \vert +2k}$, $k\in \mathbb{N}_0$. Since $\psi$ is
smooth,
its harmonic expansion is rapidly convergent, and we can conclude that
each Fourier mode $\psi_l$ behaves near $\rad=0$ as
\begin{equation}
\psi_l (\rad) = \rad^{\vert l \vert} (\psi_l^{(0)} + \rad^2
\psi_l^{(1)}
+ \dots ) .
\end{equation}

Let us see in detail what happens near $\rad_1=0$. For that we have to
look at the sum over $q_1=0$, $q_2\ne 0$. 
The corresponding terms in (\ref{1856}) read
\begin{equation}
\frac c{t^2} \int d^2 \rad\,\rad_1 \rad_2  \sum_{q_2\ne 0}
f_{q_2}(\rad)\, e^{ 
-\rad_1^2 (1-\cos \theta q_2)/2t}, 
\label{1822}
\end{equation}
where $f_{q_2}(\rad):= l_{0,q_2}(\rad) \, r_{0,-q_2}(\rad)$.
Fix $\epsilon_1 > 0$. It is easy to see that the integral 
$\int_{\epsilon_1}^\infty d\rad_1$ gives an
exponentially small term. For $\rad_1 \le \epsilon_1$ we use the
Taylor expansion of 
\begin{equation*}
f_{0,q_2}(\rad)=f^{(0)}_{0,q_2}(\rad_2)+\rad_1^2
f^{(1)}_{0,q_2}(\rad_2)+\dots
\end{equation*}

Then (\ref{1822}) takes the form (up to higher order terms):
\begin{equation}
\frac c{t^2} \frac 12 \int d\rad_2\,\rad_2 \sum_{q_2\ne 0}
\left( f^{(0)}_{0,q_2}(\rad_2) \frac {2t}{1-\cos \theta q_2} +
f^{(1)}_{0,q_2}(\rad_2)
\frac {4t^2}{(1-\cos \theta q_2)^2} +\dots \right)
\end{equation}
The sum and the integral are convergent if $\theta$ satisfies a
Diophantine condition. Recall the control of $(1-\cos\th 
q_{i})^{-1}$ by (\ref{Diophcond}). 
It is interesting to note that already the $1/t$ term receives a
contribution from the singularity.

Next, let $q_1\ne 0$, $q_2\ne 0$. The contributions from the integrals
$\int_{\epsilon_1}^\infty d\rad_1$ and $\int_{\epsilon_2}^\infty
d\rad_2$
are exponentially small. Therefore, we restrict ourselves to the
integral $\int_0^{\epsilon_1} \int_0^{\epsilon_2}d\rad_1 d\rad_2$
where we use again the Taylor expansion in the radii. The Taylor
expansion of any smooth function starts with 
$\rad_1^{q_1}\rad_2^{q_2}$ and the Taylor expansion for 
$l_{q} \cdot r_{-q}$ starts
with $\rad_1^{2q_1}\rad_2^{2q_2}$. The corresponding terms contribute
to the heat kernel coefficients with
\begin{equation}
\frac 1{t^2} \left( \frac {2t}{1-\cos \theta q_2} \right)^{q_1+1}
\left( \frac {2t}{1-\cos \theta q_1} \right)^{q_2+1}
\end{equation}
so that the modifications start with $t^2$. One can easily evaluate
corresponding terms (which describe the effect of the singularity
at $\rad_1 = \rad_2 =0$). 

The asymptotic we obtain 
strongly depends on the functions $l$ and $r$ through their
Taylor coefficients in a neighborhood of singular points.
This is typical for the heat trace asymptotics if boundaries or
singularities are present, cf.~\cite{newGilkey,Vassilevich:2003xt}.

\section{Spectral action for NC-tori}
\label{spectralaction}

Within noncommutative geometry, the spectral action introduced by 
Chamseddine--Connes plays an important role \cite{CC}. More 
precisely, given a spectral triple $(\A,\H,\DD)$ where $\A$ is an 
algebra acting on the Hilbert space $\H$ and $\DD$ is a Dirac-like
operator (see \cite{Book,Polaris}), they proposed a physical action 
depending only on the spectrum of the covariant Dirac operator
\begin{equation}
\label{covDirac}
\DD_{A}:=\DD + A + \epsilon \,JAJ^{-1}
\end{equation}
where $A$ is a one-form represented on $\H$, i.e. it is of the form 
\begin{equation}
\label{oneform}
A=\sum_{i}a_{i}[\DD,\,b_{i}],
\end{equation}
where $a_{i}$, $b_{i}\in \A$, $J$ is a real 
structure on the triple corresponding to charge conjugation and 
$\epsilon \in \set{1,-1}$ depending on the dimension of this triple 
and comes from the commutation relation 
\begin{equation}
    \label{Jcom}
J\DD=\epsilon \, \DD J.
\end{equation}

This action is
\begin{equation}
\label{action}
\SS(\DD_{A},\Phi):=\Tr \big( \Phi(\DD_{A}^2/\Lambda^2) \big)
\end{equation}
where $\Phi$ is any positive function viewed as a cut-off which could 
be replaced by a step function up to some mathematical difficulties 
surmounted in \cite{Odysseus}. This means that $\Phi$ counts the 
spectral values of $\vert \DD_{A} \vert$ less than the mass scale 
$\Lambda$ (note that the resolvent of $\DD_{A}$ is compact since, by 
assumption, the same is true for $\DD$).

In \cite{GI2002}, the spectral action on NC-tori has been computed 
only for operators of the form $\DD + A$. Thanks to our previous 
result, we can fill the gap and compute (\ref{action}) in full 
generality.

We need to fix notations: Let $\A_{\Th}:=C^{\infty}(\T_{\Th}^n)$ 
acting on $\H:=\H_{\tau}\otimes \C^{2^m}$ with $n=2m$ or $n=2m+1$ 
(i.e., $m=\lfloor \tfrac n2 \rfloor$ is the integer part of $\tfrac
n2$),
the square integrable sections of the trivial spin bundle over $\T^n$.

Each element of $\A_{\Th}$ is represented on $\H$ as 
$L(a)\otimes1_{2^m}$. The Tomita conjugation $$J_{0}(a):=a^*$$ 
satisfies $[J_{0},\pa_{\mu}]=0$ since
$J_0\pa_{\mu} U_k=J_0(ik_\mu)U_{k}=-ik_\mu U_{-k}=
\pa_\mu U_{-k}=\pa_\mu J_0 U_k.$
Besides, it induces the analogous operator on $\H$, 
$$
J:=J_{0}\otimes C_{0}
$$
where $C_{0}$ is an operator on $\C^{2^m}$.
The Dirac operator is defined by
$$
\DD:=-i\, e_{a}^{\mu}\,\delta_{\mu} \otimes \gamma^{\a}=-i\, 
\delta_{\mu}\otimes \gamma^{\mu},
$$
where we use hermitian Dirac matrices $\gamma$. This implies 
$C_{0}\ga^{\mu}=-\eps \ga^\mu C_{0}$ since 
$$
J\DD=(J_0\ox C_0)(-i\pa_\mu\ox\ga^\mu)=J_0(-i)\pa_\mu\ox C_0\ga^\mu
=i\pa_\mu J_0\ox \ga^\mu C_{0},
$$
which by (\ref{Jcom}) is equal to
$\eps (-i\pa_\mu)J_0\ox \ga^\mu C_0$.
Moreover, $C_{0}^2=\pm 1_{2^m}$ depending on the parity of $m$.
Finally, one introduces the chirality (which in the even 
case is $\chi:=id \otimes (-i)^{m} \gamma^1 \cdots \gamma^{n}$) and 
this yields that $(\A_{\Th},\H,\DD,J,\chi)$ 
satisfies all axioms of a spectral triple, see \cite{Book,Polaris}.

The unitary elements $u$ of $\A_{\Th}$ (or of its generated
C$^*$-algebra) 
play an important role since they reflect the inner automorphisms of 
$\A_{\Th}$. For instance $U_{u}:=(u\otimes 1_{2^m})J(u\otimes 
1_{2^m})J^{-1}$ is a unitary on $\H$ (with  
$U_{u}^*=(u^*\otimes 1_{2^m})J(u^*\otimes 1_{2^m})J^{-1}$)
such that
$$
U_{u}\DD U_{u}^*= \DD + u\otimes 1_{2^m}[\DD,u^*\otimes 1_{2^m}]+ 
\eps J u\otimes 1_{2^m}[\DD,u^*\otimes 1_{2^m}]J^{-1}
$$
explaining the construction of one-forms $A$ in (\ref{oneform}) 
thus satisfying
$U_{u}AU_{u}^*=u\otimes 1_{2^m}Au^*\otimes 1_{2^m}$. 
These properties follow from the axioms: for all $a,b \in 
\A_{\Th}$, 
$[a\otimes 1_{2^m},Jb\otimes 1_{2^m}J^{-1}]=0$ and $
\big[[\DD,a\otimes 
1_{2^m}],Jb\otimes 1_{2^m}J^{-1}\big]=0 .$

In conclusion, the fact that the perturbed Dirac operator must satisfy 
condition (\ref{Jcom}) (which is equivalent to $\H$ being
endowed with a structure of $\A_{\Th}$-bimodule: for $a,b\in\A_{\Th}$ and 
$\psi \in \H_{\tau}$, 
$a.\psi.b:=a\otimes 1_{2^m} 
Jb^*\otimes 1_{2^m}J^{-1} \, \psi$), yields the necessity of a
symmetrized covariant Dirac operator: $\DD_{A}=\DD + A + \epsilon 
J\,A\,J^{-1}$.

Note that for $a \in \A_{\Th}$, using $J_{0}L(a)J_{0}^{-1}=R(a^*)$,
\begin{align}
\label{LR}
\epsilon J\big(L(a)\otimes 
\gamma^{\mu}\big)J^{-1}&= \eps \, R(a^*) 
\otimes C_{0}\ga^\mu C_{0}^{-1}=-R(a^*)\otimes \gamma^{\mu},
\end{align}
and that the representation $L$ and the 
antirepresentation $R$ are $\C$-linear, commute and satisfy  
$$[\delta_{\mu},L(a)]=L(\delta_{\mu}a),\quad
[\delta_{\mu},R(a)]=R(\delta_{\mu}a).$$

Choosing an arbitrary selfadjoint one-form $A$, it 
can be written as
\begin{equation}
\label{connection}
A = L(-iA_{\mu})\otimes\gamma^{\mu},\,\, A_{\mu} =-A_{\mu}^* \in 
\A_{\Th}
\end{equation} 
and using (\ref{LR})
\begin{equation}
\label{dirac}
\DD_{A}=-i\,\big(\delta_{\mu}+L(A_{\mu})-R(A_{\mu})\big)
 \otimes \gamma^{\mu}.
\end{equation}
Defining $$\tilde A_{\mu}:=L(A_{\mu})-R(A_{\mu}),$$
we get
$$
\DD_{A}^2=-g^{\mu \nu}(\delta_{\mu}+\tilde 
A_{\mu})(\delta_{\nu}+\tilde A_{\nu})\otimes 1_{2^m} - \tfrac 12 
\Omega_{\mu \nu}\otimes \gamma^{\mu \nu}
$$
where $\gamma^{\mu \nu}:=\tfrac 12 
(\gamma^{\mu}\gamma^{\nu}-\gamma^{\nu}\gamma^{\mu})$ and  
\begin{align*}
\Omega_{\mu \nu}&:=[\delta_{\mu}+\tilde A_{\mu},\delta_{\nu}+\tilde 
A_{\nu}] \\
&\,=L(F_{\mu \nu}) - R(F_{\mu \nu})
\end{align*}
where
$$
F_{\mu \nu}:=\delta_{\mu}(A_{\nu}) 
-\delta_{\nu}(A_{\mu})+[A_{\mu},A_{\nu}].
$$
Gathering all results,
\begin{align}
\label{D2}
\DD_{A}^2=-g^{\mu \nu} \Big( \delta_{\mu}+L(A_{\mu})-R(A_{\mu})\Big)
\Big(\delta_{\nu}+L(A_{\nu})-R(A_{\nu})\Big) \otimes 1_{2^m}
\nonumber\\
-\tfrac 12\,\big(L(F_{\mu \nu}) - R(F_{\mu \nu})\big) 
\otimes \gamma^{\mu \nu}
\end{align}
Now, comparing (\ref{D2}) and (\ref{operD}), we can apply the previous 
result with the following replacement in (\ref{covdir}) and 
(\ref{defE}) 

\begin{align}
\label{replacemen}
&\begin{cases}
L(\lambda_\mu)  & \rightarrow \;L(\,A_\mu)\otimes 
1_{2^m}  ,\\
R(\rho_{\mu}) &\rightarrow \; R(\, A_\mu)
\otimes 1_{2^m}  ,\\
L(l_{1}) &\rightarrow  \; -\tfrac 12  L(F_{\mu \nu})\otimes 
\gamma^{\mu \nu}  ,\\
R(r_{1}) &\rightarrow  \; -\tfrac 12 R(F_{\mu \nu})\otimes 
\gamma^{\mu \nu}  ,\\
L(l_{2}) &\rightarrow  \; 0, \\
R(r_{2}) &\rightarrow  \; 0.
\end{cases}
\\
\noalign{\noindent \text{or in Theorem \ref{theorem1}}}
\label{replacemenbis}
&\begin{cases}
\nabla_{\mu}  & \rightarrow \;\big(\delta_{\mu}
+L(A_\mu)-R(A_{\mu})\big) \otimes 1_{2^m}  ,\\
E &\rightarrow  \; -\tfrac 12  \big( L(F_{\mu \nu})
- R( F_{\mu \nu})\big) \otimes 
\gamma^{\mu \nu} ,\\
\widehat{\Omega}_{\mu \nu}  & \rightarrow \;\big(L(F_{\mu \nu}) 
-R(F_{\mu \nu})\big)\otimes 1_{2^m}, \\
L(l) & \rightarrow \; 1,\\
R(r) & \rightarrow \; 1.\\
\end{cases}
\end{align}

It is interesting to note that, in the commutative case when 
$\Th=0$, $L=R$ thus $\DD_{A}=\DD$ for any selfadjoint one-form $A$: 
the Dirac operator does not fluctuate.

We will derive the spectral action by Laplace transform 
techniques such as in \cite{Nestetal}, see \cite{Widder} for
details on Laplace transform (alternatively one can follow 
\cite{Odysseus}).
We assume that the function $\Phi$ has the 
following property:
\begin{equation}
\label{condition}
\Phi\in\mathcal{C}^\infty(\R^+) \!\!\! \sepword 
{is the Laplace transform of} \!\!\!
\hat{\psi} \in \SS(\R^+):=\set{g\in\SS (\R): g(x)=0, x \leq 0}
\end{equation}
Thus, any function with this property has 
necessarily an analytic extension on the right complex 
plane and is a Laplace transform.
Consequently, any $m$-differentiable function 
$\psi$ such that $\psi^{(m)}=\Phi$ is the
Laplace transform of a function $\hat{\psi}$ and 
by differentiation, it satisfies
\begin{equation*}
 \Phi(z)=\psi^{(m)}(z)=(-1)^m\int_0^\infty e^{-tz}\; 
 t^m\;\hat{\psi}(t)\;dt,\;\Re z>0.
\end{equation*}
One can invoke dominated convergence (see \cite{GI2002}), to obtain:
\begin{eqnarray*}
\Tr\left(\Phi(\DD_{A}^2/\Lambda^2)\right) 
&=&(-1)^m\int_0^\infty\Tr\left(e^{-t\,
\DD_{A}^2/\Lambda^2}\right)t^m \, \hat{\psi}(t) \; dt\\
&=&(-1)^m \int_0^\infty\sum_{k=0}^{m}
\Lambda^{n-2k}\;\tilde{a}_{2k} \, t^{m+k-n/2} \,
\hat{\psi}(t) \; dt\; +\;\mathcal{O}(\Lambda^{n-2(m+1)})\\
&=&\sum_{k=0}^{m}\Lambda^{n-2k}\,\Phi_{2k}
\, \tilde{a}_{2k}\;+\;\mathcal{O}(\Lambda^{n-2(m+1)}),\\
\end{eqnarray*}
where $\Phi_{2k}$ is defined by
\begin{equation}
\label{moments2}
\Phi_{2k}:=(-1)^m\int_0^\infty t^{m+k-n/2}\,\hat{\psi}(t) \; dt,
\end{equation}
and
\begin{equation}
\label{atilde}
\tilde{a}_{2k}:=a_{2k}(1,1;\DD_{A}^2)
\end{equation}
obtained from (\ref{aoriginal}) with replacement (\ref{replacemen}).

When $n=2m$ is even, $\Phi_{2k}$ has the more familiar form:
\begin{eqnarray}
\label{moments3}
\Phi_{2k}=
\begin{cases}
{\textstyle{\frac{1}{\Gamma(m-k)}}}\;\int_0^\infty \Phi(t) 
\; t^{m-1-k} \; dt,\;& {\rm for}\;
k=0,\cdots,m-1,\cr
(-1)^{k}\;\Phi^{(k-m)}(0),& {\rm for}\; k=m,\cdots,n.
\end{cases}
\end{eqnarray}
For $n$ odd, the coefficients
$\Phi_{2k}$ have less explicit forms because 
they involve fractional derivatives of $\Phi$,
so in this case, it is better to stick with 
definition (\ref{moments2}).

Let us summarize:

\begin{theorem}
Let $\DD_{A}=\DD+A + \epsilon JAJ^{-1}$ and $A = L(-iA_{\mu})\otimes 
\gamma^{\mu}$ a hermitian one-form where 
$A_\mu^*=-A_\mu\in \A_{\Th}=C^{\infty}(\T^n_{\Th})$. Let $\Phi
\in\mathcal{C}^\infty(\R^+)$
be a positive function satisfying condition 
(\ref{condition}). Then the following expansion of the 
spectral action holds:
$$
\SS(\DD_{A},\Phi)=
 \sum_{k=0}^{\lfloor n/2 \rfloor}\Lambda^{n-2k}
\; \Phi_{2k}\; \tilde{a}_{2k}\;+\;\mathcal{O}(\Lambda^{n-2(\lfloor n/2
\rfloor+1)}),
$$
where the $\Phi_{2k}$ are defined in (\ref{moments2}) or
(\ref{moments3}) depending on the dimension and the
$\tilde{a}_{2k}$  are defined in (\ref{atilde})
with replacement (\ref{replacemen}).

More precisely,
\begin{align*}
\tilde{a}_{0} & = 2^{\lfloor n/2 \rfloor} \pi^{n/2},\\
\tilde{a}_{2} & =0 ,\\
\tilde{a}_{4} & = (4\pi)^{-n/2}\, \left[ \tfrac 12 \Sp (E^2) +
\tfrac 1{12}\Sp (\widehat{\Omega}^{\mu \nu} \,
\widehat{\Omega}_{\mu \nu}) \right] .
\end{align*}

\noindent Moreover, all terms in $\tilde a_{2k}$ linear in 
$l_{1},r_{1}$ are zero.
\end{theorem}
\begin{proof}
The last assertions follow from $\tilde{a}_{0} =
(4\pi)^{-n/2}\,
\Sp (1\otimes 1_{2^m})$ , $\tilde{a}_{2} =(4\pi)^{-n/2}\, \Sp (E)$, 
$\tilde{a}_{4} 
= (4\pi)^{-n/2}\, \left[ \tfrac 12 \Sp (E^2) + \tfrac 16 \Sp (g^{\mu 
\nu}\,E_{;\mu \nu}) + \tfrac 1{12}\Sp (\widehat{\Omega}^{\mu \nu} \, 
\widehat{\Omega}_{\mu \nu})\right]$ and $\Tr(\gamma^{\nu \mu})=0$,
so all linear terms in $E$ are of zero trace.
\end{proof}

Now comes an amazing fact: in four dimensions, the non-standard terms
(those with products of traces) simply disappear. Indeed, when $n=4$,
with $g=\diag(1,1,1,1)$ and $\Th$ a direct
sum of two ``Diophantine'' matrices, we find, up to negative powers of
$\Lambda$,
\begin{align*}
\SS(\DD_{A},\Phi)&=(4\pi)^{-n/2}4\Big(\Phi_0\Lambda^4 \Sp(1)
-\tfrac{\Phi(0)}{6}
\Sp\big(L(F^{\mu\nu}F_{\mu\nu})+R(F^{\mu\nu}F_{\mu\nu})
-2L(F^{\mu\nu})\,R(F_{\mu\nu})\big)\Big)\\
&=4\pi^{n/2}\Big(\Phi_0\Lambda^4-\tfrac{\Phi(0)}{3}\big(
\tau(F^{\mu\nu}F_{\mu\nu})-\tau(F^{\mu\nu})\,\tau(F_{\mu\nu})\big)\Big).
\end{align*}
But $F^{\mu\nu}$ is a sum of derivatives plus a commutator, so is
of zero trace. Thus, the spectral action has the standard form:
\begin{align}
\label{spectralaction1}
\SS(\DD_{A},\Phi)=4\pi^{n/2}\Big(\Phi_0\Lambda^4-\tfrac{\Phi(0)}{3} \,
\tau(F^{\mu\nu}F_{\mu\nu})\Big)+\;\mathcal{O}(\Lambda^{-2}).
\end{align}
The only difference appears in the numerical value of the coefficients.

What happens for generic compact toric NC manifolds (we add the
assumption that $M$ carries a spin structure as well)?

Even lacking a trace asymptotic expansion of the semi-group
generated by a generalized Laplacian (i.e., an analogue of theorem
\ref{theorem1}), we are able to finish in the 4-d pure Diophantine
case, using the symmetries we have at disposal. First, it should be
clear from examples treated in previous section that the
supplementary terms coming from the `singular points' actually do not
appear. Indeed, they should appear only in the sub-leading order of a
given term, but here the only one we have is the `Yang--Mills' one,
which is already the last with non-negative power of $\Lambda$ (i.e.,
such correction terms appears in 4-d with negative powers of
$\Lambda$). \\
{\it In summary, the torus action being free or not has no serious
implication for the structure of the spectral action in dimension four
or less}. One should emphasize that this is no longer true in higher
dimensions. 

What previous examples also do show is that, up to
sub-leading order terms,
$$
\Tr\big(L(l)\,R(r)\,e^{-t\DD^2}\big)=\int_M\mu_g(p)\,l_0(p)\,r_0(p)\,
\widetilde{K}_t(p,p),
$$
where $\widetilde{K}_t$ is the on-diagonal kernel of
$e^{-t\DD^2}$. However, $a_0=0$ whenever $a$ is either a commutator
or a derivative (with respect to the action $\a$) in $\Coo(M_\Th)$.
Thus
the same conclusion holds, namely, that
the asymptotics of
$\Tr[\Phi\big((\DD+A+\epsilon JAJ^{-1})^2/\Lambda^2\big)]$ can be
easily derived from those of
$\Tr[\Phi\big((\DD+A)^2/\Lambda^2\big)]$, in 4-d 
with Diophantine deformation matrix. Note that the latter is
easily computable from the classical asymptotics of the kernel 
$\widetilde{K}_t(p,p)$. 

To compute the spectral action, it can also be convenient to 
use the full force of \cite{CC1}.

\section{NC-QFT: Structure of divergences for a scalar 
field theory}
\label{sec-ncqft}

Let us
consider a real scalar field $\phi$ in a four-dimensional NC torus 
with the classical action
\begin{equation}
S[\phi]=\int d^4x \sqrt{g} \left( \tfrac 12 \,(\partial_\mu \phi)^2 +
\tfrac 12 m^2 \phi^2 + \tfrac {\lambda}{24}\, \phi\star\phi\star\phi\star\phi
\right),
\label{p4act}
\end{equation}
where ${\lambda}$ is a coupling constant. Here we change to the notations
which are
more common in quantum field theory and write the star products
(\ref{Rieffel}) or (\ref{StarP}) and partial derivatives.
To calculate the effective action in this theory, we split 
$\phi =\varphi - \delta\phi$
into a background part $\varphi$ and quantum fluctuations
$\delta\phi$.
Then one
expands $S[\phi]$ about the background. The first term, $S[\varphi]$,
simply gives the classical approximation to the effective action,
The second term, which is proportional to the first derivative of 
$S[\varphi]$ is canceled by external sources. The quadratic term
can be rewritten as
\begin{equation*}
S^{(2)}[\vf,\delta\phi]=
\tfrac 12 \int d^4x\,\sqrt{g}\,(\delta\phi) P (\delta \phi) \,,
\end{equation*}
where (cf.~\cite{Gayral:2004cu})
\begin{equation}
P=-g^{\mu\nu}\partial_\mu\partial_\nu 
+ \tfrac {\lambda}6 \left[ R(\varphi \star \varphi )+
L(\varphi \star \varphi )+L(\varphi )\,R (\varphi )\right]+m^2\,.
\label{Dp4}
\end{equation}
Note that $P>0$ for $\lambda >0$,
since $-g^{\mu\nu}\partial_\mu\partial_\nu \geq0$ and $\vf^*=\vf$ so that
$$
R(\varphi \star \varphi)+
L(\varphi \star \varphi )+L(\varphi )\, R (\varphi )=\tfrac12
\big(L(\vf)+R(\vf)\big)^*\big(L(\vf)+R(\vf)\big)+ \tfrac12
L(\vf)^*L(\vf) +\tfrac12 R(\vf)^*R(\vf).
$$
This operator corresponds to the following choice in 
(\ref{operD}--\ref{defE})
\begin{equation*}
\lambda_\mu =\rho_\mu =0\,,\quad 
l_1=-r_1=-\tfrac {\lambda}6 \,\varphi \star \varphi -\tfrac {m^2}2
\,,\quad l_2=-r_2=\sqrt{\tfrac {\lambda}6} \,\varphi \,.
\end{equation*}
Formally, the one-loop effective action reads
\begin{equation*}
W=\tfrac 12 \ln \det P \,.\end{equation*}
This expression has to be regularized. We shall use the zeta-function
regularization \cite{Dowker,Hawking}. 
The zeta function can then be defined as an $L^2$-trace
\begin{equation*}
\zeta (s,P) = \Tr ( P^{-s}) \,.
\end{equation*}
The pole structure of the zeta function is determined by the 
asymptotic properties of  heat trace at $t\to 0$ 
(see, e.g. \cite{newGilkey}). Due to
Theorem \ref{theorem1}, the zeta function of $P$ has only simple poles 
and is regular at $s=0$. There is a useful relation
\begin{equation*}
a_k(P)={\mathrm {Res}}_{s=\frac{n-k}2} \big(\Gamma (s) \,\zeta 
(s,P)\big).
\end{equation*}
In particular, $a_n=\zeta (0,P)$. 

The regularized effective action is defined as
\begin{equation*}
W(s)=-\tfrac 12 \,\mu^{2s} \Gamma (s) \,\zeta (s,P) \,.
\end{equation*}
The regularization is removed in the limit $s\to 0$. $\mu$ is a 
dimensional constant introduced to keep the regularized effective
action dimensionless. Because of a pole of the $\Gamma$-function,
$W_s$ is divergent at $s\to 0$.
The divergent part of the effective action reads
\begin{equation*}
W_{\rm div}(s)=-\tfrac 1{2s} \,\zeta (0,P)=-\tfrac 1{2s} \,a_4
(P).
\end{equation*}
Let us assume that $\Theta$ satisfies a
Diophantine condition. Then, by eq.\ (\ref{a4lr}),
\begin{eqnarray}
&&\hspace{-1cm}a_4(P)=\frac 1{32\pi^2} \,\left[ v m^4 -\frac{{\lambda}m^2}3 
\big( 2 \int d^4x \,\sqrt{g}\, \varphi^2 + v^{-1} ( 
\int d^4x\sqrt{g} \,\varphi )^2 \,\big) \right. \nonumber\\
&&\quad \left. + \frac{{\lambda}^2}{36} \,\big( 2 \int d^4x \sqrt{g}\,
\varphi_\star^4
+ 3 v^{-1} ( \int d^4x \sqrt{g}\,\varphi^2 )^2
+ 4 v^{-1}\int d^4x\sqrt{g} \,\varphi_\star^3
\int d^4x \sqrt{g}\,\varphi \, \big) \right]. \label{a4phi4}
\end{eqnarray}
Here $v={\rm vol}\, \T^4=(2\pi)^4 \sqrt{g}$,
$\varphi_\star^k$ denotes the $k$-th star-power of $\varphi$. For example,
$\varphi_\star^3:=\varphi \star \varphi \star \varphi$.

The theory is called form-renormalizable if all divergences in the
effective action can be compensated by redefinitions of couplings
in the classical action, i.e., if the divergent part of the effective
action repeats the structure of the classical action. The term $m^4$
in (\ref{a4phi4}) causes no problem as it does not depend on the field
(it is said that such terms are removed by a renormalization of the
cosmological constant). The terms with $m^2\varphi^2$ and
$\varphi_\star^4$
can be removed by suitable renormalization of $m^2$ and ${\lambda}$ in
(\ref{p4act}).
The remaining non-local terms cannot be renormalized away. Therefore,
the model (\ref{p4act}) is {\em not} form-renormalizable.

It is instructive to consider an infinite-volume limit of (\ref{a4phi4}).
Let us introduce normalized coordinates $y^a=e_\mu^a x^\mu$, where
$e^a_\mu$ is a constant vierbein, $e^a_\mu e^a_\nu =g_{\mu\nu}$,
and let us assume that $\varphi (y^a)$ is kept constant as $v\to\infty$.
Then all nonlocal terms vanish in the limit $v\to\infty$. This is
consistent with the conclusion of \cite{Vassilevich:2005vk} that 
the counterterms for $\phi^4$ on $\mathbb{R}^4_\Th$
are local in the zeta function regularization if $\Th$ is nondegenerate.
Note that, for a degenerate $\Th$, this is no longer true 
\cite{Gayral:2004cu}. This is because the IR divergence developed in
the non-planar sector of the 2-point function (proportional to
$|\Th\xi|^{-2}$ in momentum space), turns out to be non-locally
integrable and thus the associated Green function does not define a
temperate distribution.

Of course, one can add several terms to the classical action which
repeat
the structure of the one-loop divergences. However, such terms
change,
in general, also the divergent part of the effective action and bring
up
new structures. Can this process be closed after several steps? Below
we show that, at least in one-loop approximation, the answer is
affirmative.

First of all, we impose antiperiodic conditions in one of the
coordinates, say $x^1$ on the field 
$\phi$ (and also on the background field $\varphi$, and on
quantum fluctuations $\delta \phi$): 
\begin{equation}
\phi (x^1,x^2,x^3,x^4)=-\phi (x^1 +2\pi,x^2,x^3,x^4).
\label{antiper}
\end{equation}
In the language of NC-QFT, this corresponds to considering a field theory on
a finite projective module over the noncommutative torus. This
Hermitian module is well defined. One first lifts the torus action on
the commutative vector bundle and then defines the module structure on
the smooth sections via a star product made out of the
lifted torus action (see \cite{ConnesDV}).

This anti-periodic condition cancels all terms with 
linear integral of 
the field, $\int d^4x\, \varphi=0$. By condition
(\ref{antiper})
also the Laplacian spectrum is changed, $\Delta \sim p_\mu
p^\mu$,
$p\in \mathbb{Z}^n_{(1/2)}:=\mathbb{Z}^n+(1/2,0,0,0)$. However,
as $t\to 0$, 
\begin{equation*}
\sum_{k\in \mathbb{Z}} e^{-(k+1/2)^2t}= \sqrt{ \tfrac{\pi}t} + \mbox{ e.s.t.}
\end{equation*}
and all asymptotic relations derived above remain true after obvious
modifications. Since quantum fluctuations are anti-periodic, one has
to
take in (\ref{1847}) $k\in \mathbb{Z}^n_{(1/2)}$. The fields $l$ and
$r$
are powers of the background field $\varphi$. Therefore, $r$ and $l$
may be both periodic and anti-periodic. Thus, $q\in \mathbb{Z}^n \cup
\mathbb{Z}^n_{(1/2)}$ in that equation (but only half of the Fourier
coefficients may be non-zero in each particular case). The
Diophantine condition also should be understood with respect to
$k\in \mathbb{Z}^n_{(1/2)}$ and $q\in \mathbb{Z}^n \cup
\mathbb{Z}^n_{(1/2)}$.

To deal with the remaining non-local
terms in (\ref{a4phi4}) we add to the classical action a non-local
part 
\begin{equation}
\delta S[\phi]= \frac{\tilde{\lambda}}{2\sqrt{g}}  
\left[ \int d^4x \sqrt{g} \,\phi^2
\right]^2 ,
\label{delS}
\end{equation}
where $\tilde{\lambda}$ is a new coupling constant. Renormalization of
$\tilde\lambda$ allows to remove all existing one-loop divergences.
It is important to make sure that no new types of divergences
appear. Because of (\ref{delS}), the quadratic form of the action
receives the contribution
\begin{equation}
\delta S^{(2)}[\vf,\delta\phi]= \frac{2\tilde{\lambda}}{ \sqrt{g}}
\left[ \int d^4x\sqrt{g} \,
\varphi\cdot (\delta\phi) \right]^2
+ \frac{\tilde{\lambda}}{\sqrt{g}}
  \int d^4x\sqrt{g} \, \varphi^2 \,\,
\int d^4x \sqrt{g}\, (\delta\phi)^2\,.
\label{delS2}
\end{equation}
The second term in (\ref{delS2}) is harmless. Due to that, the term $m^2$
in  (\ref{Dp4}) is replaced by a background-dependent but still
coordinate-independent mass term:
\begin{equation}
m^2\to \bar m^2 = m^2 + 2 \tilde{\lambda}  \int d^4x \, \varphi^2.
\label{barm2}
\end{equation}
Let us assume $\tilde\lambda >0$ so that after this 
replacement the spectrum of $P$
remains positive. 
Next we substitute $\bar m^2$ for $m^2$ in (\ref{a4phi4}) and take
into account that $\int d^4x \, \varphi =0$ to see that the divergent
part of the effective action does not receive any new structure
beyond those which are already present in $S+\delta S$. 

The first term on the right hand side of (\ref{delS2}) does not
contribute to the divergences at all. Let us denote by $\bar P$
the operator (\ref{Dp4}) with the replacement (\ref{barm2}). Then, the
operator acting on quantum fluctuations reads
\begin{equation*}
P=\bar P + B,\qquad 
B=4 {\tilde{\lambda}} g^{-1/2} \,\vert \varphi\rangle  \langle \varphi \vert 
\end{equation*}
where $\vert \varphi\rangle  \langle\varphi\vert$ is a rank one
operator
(with suggestive notations). 
$B$ being proportional to a projector 
to a one-dimensional subspace of the Hilbert space,
it is clear that the insertions of $B$ improve ultraviolet behavior
of quantum amplitudes. To make this argument more precise we use the
Duhamel principle (\ref{Duhamel}):
\begin{equation*}
e^{-t(\bar P+B)} = \sum_{j=0}^\infty (-t)^j \beta_j \,.
\end{equation*}
For $j\ge 1$,
\begin{eqnarray*}
\Tr \beta_j &=& \int_{\Delta_j} d^js \, \Tr \left(
Be^{-(s_2-s_1)t\bar P} B \dots e^{-(1-s_j+s_1) t\bar P}
\right)\nonumber\\
&=& \frac{(4\tilde{\lambda})^j}{g^{j/2}} 
 \int_{\Delta_j} d^js \, \langle \varphi \vert e^{-(s_2-s_1)t\bar P}
\vert \varphi \rangle \dots
\langle \varphi \vert e^{-(1-s_j+s_1)t\bar P}
\vert \varphi \rangle \,.
\end{eqnarray*}
Since under our assumptions the operator $\bar P$ is positive,
$|\langle \varphi \vert e^{-s\bar P}
\vert \varphi \rangle |\le \| \varphi \|^2_{L^2}$ for $s\ge 0$. Thus
\begin{equation*}
\vert \Tr \beta_j \vert \le \frac{ \vert \tilde{4\lambda} \vert^j 
  \| \varphi \|^{2j}_{L^2}}{j!g^{j/2}},
\end{equation*}
and this means that the series expansion
\begin{equation*}
\Tr (e^{-t(\bar P +B)}) -\Tr (e^{-t\bar P})=
\sum_{j=1}^\infty (-t)^j \Tr \beta_j 
\end{equation*}
is absolutely convergent in the UV regime, and
\begin{equation*}
\Tr (e^{-t(\bar P +B)}) -\Tr (e^{-t\bar P})=\mathcal{O}(t)\,,
\end{equation*}
i.e., the operator $B$ does not contribute to the coefficient
$a_4(\bar P +B)$ or to the one-loop divergences.

The action $S+\delta S$ with anti-periodic boundary conditions
is indeed renormalizable at one loop.
The canonical mass dimension of the coupling constant $\tilde e$
is $+4$. Standard power-counting arguments show that the insertions
of the interactions with $\tilde e$ can only improve the ultraviolet
behavior of corresponding Feynman diagrams (since in the momentum
cut-off regularization each positive power
of $\tilde e$ should be accompanied by negative powers of the cut-off
momentum). Although this power-counting may break down for
noncommutative theories, it is nevertheless natural to conjecture 
that the theory
with $S+\delta S$ will remain renormalizable also at higher orders of
the loop expansion.

The need to add non-standard terms to the NC action in order to
achieve renormalizability is not really surprising (see
\cite{Grosse:2004yu}). Note that in the approach of
\cite{Grosse:2004yu}, the Diophantine condition does not play any
role. The difference probably comes from the fact that we are
working with compact noncommutative directions right from the
beginning, while in \cite{Grosse:2004yu} the "compactification"
appears dynamically due to the presence of an oscillator
potential. 
Physical consequences of adding the non-local term (\ref{delS}) to
the action are still unclear to us. Since we consider the case of a
compact
Euclidean manifold (a torus) there are no immediate problems with 
causality
(note that on $\mathbb{R}^n_\Theta$, no terms like (\ref{delS}) are
required for the one-loop renormalization \cite{Vassilevich:2005vk}). 
Alternatively, noncommutative theories may be viewed as effective
low-energy
theories, so that renormalizability is not required. In this case
one
needs a self-consistent subtraction scheme only. An example of such a
scheme is
given by the large-mass subtraction \cite{Bordag:1998vs}, which also
uses the asymptotic expansion of the heat trace.

The technical tools developed in the previous sections are
sufficient to analyze other fields (spinors, gauge fields, etc).
In this paper, we restricted ourselves to abelian gauge fields. An
extension to non-abelian gauge fields can be done rather
straightforwardly. Probably, even an extension to superfields
can be achieved since the technique used in superspace 
\cite{Azorkina:2006te} is similar to the one presented here.

\appendix{\setcounter{section}{-1}}
\section{Appendix}

\section{Proof of asymptotic formulae}\label{appA}

Here we prove equation (\ref{asym1}). Starting from \eqref{1847}
\begin{equation*}
 \Tr \left( L(l)\, R(r)\, e^{-t\Delta}\right) = \sqrt g\,(4\pi
t)^{-\frac n2}
\sum_{q\in \mathbb{Z}^n} \sum_{k\in \mathbb{Z}^n} 
l_{q}\, r_{-q} \,\, e^{ -\tfrac{|\Theta q -2\pi k|^2}{4t}},
\end{equation*}
we split the sum over $q$ as
\begin{equation}
\sum_{q\in \mathbb{Z}^n}=
\sum_{q\in \mathcal{Z}}+\sum_{q\in \mathcal{K}}.
\label{1620}
\end{equation}
Consider the sum over $\mathcal{Z}$ first. Since for $q\in
\mathcal{Z}$
the vector $\Theta q/2\pi$ belongs to $\mathbb{Z}^n$, one can shift
$k$ in the subsequent sum by $\Theta q/2\pi$. This yields
\begin{equation*}
 \sqrt g\,
(4\pi t)^{-\frac n2} \sum_{q\in \mathcal{Z}}
\Big(  l_{q}\,r_{-q}\, +
\sum_{0 \neq k\in \mathbb{Z}^n} 
l_{q}\,r_{-q}\,  e^{-\frac{\pi^2 \vert k\vert^2}t} \Big) 
=(4\pi t)^{-\frac n2} \sum_{q\in \mathcal{Z}} l_{q} \, r_{-q} \,+
\mbox{ e.s.t.}
\end{equation*}
since
$$
\Big| \sum_{q\in \mathcal{Z}}
 l_{q}\,r_{-q}\sum_{0 \neq k\in \mathbb{Z}^n} 
e^{-\frac{\pi^2 \vert k\vert^2}t}\Big|\leq
2\,e^{-\frac{\pi^2}{t}}\sum_{q\in \mathcal{Z}}|
l_{q}\,r_{-q}|\sum_{k\in\N^n}e^{-\frac
{\pi^2(\vert k\vert^2+2k)}{t}}\leq C \, e^{-\frac{\pi^2}{t}},
$$
because $\sum_{k\in\N^n}e^{-\pi^2(\vert k\vert^2+2k)/t}$ is uniformly
bounded in $t$, and $\{l_{q}\,r_{-q}\}$ is a Schwartz sequence.

Next we have to consider the second sum in (\ref{1620}). For each
$q\in \mathcal{K}$ we choose 
$k_0(q) \in \mathbb{Z}^n$ which minimizes the distance to $\Theta 
q/(2\pi)$ (if there are several such $k$, we can take any one of them.) 
For $k\ne k_0$, one can estimate $|\Theta q -2\pi k|^2 \ge 4\pi^2 
(c_1+c_2 \vert k-k_0\vert^2)$. From now on,  $c_i$'s denote some
positive
constants. Therefore, the terms with $k\ne k_0$ give only
exponentially 
small terms in the second sum of (\ref{1620}). It remains to 
evaluate the sum
\begin{equation}
\mathcal{S}=\sqrt g  \,(4\pi t)^{-n/2}
\sum_{q\in \mathcal{K}} 
l_{q} \, r_{-q} e^ {-|\Theta q -2\pi k_0(q)|^2/4t} .
\label{1638}
\end{equation}
Note that $\Theta q -2\pi k_0(q) \ne 0$ in (\ref{1638}). We split the
sum in (\ref{1638}) into the sums over $\mathcal{K}^{\rm per}$ and
over $\mathcal{K}^{\rm inf}$. The first sum consists of a finite
number of exponentially small terms.
Consequently, it is exponentially small itself. For 
$q\in \mathcal{K}^{\rm inf}$ we use Diophantine condition 
(\ref{Diop}) to obtain
\begin{equation*}
|\mathcal{S}| \le \sqrt g\,(4\pi t)^{-n/2}
\sum_{q\in \mathcal{K}^{\rm inf}}
|l_{q} \, r_{-q} |
e^{-C^2/4 |q|^{2(1+\beta)} t} + \mbox{ e.s.t.}
\end{equation*}
Let us again divide the sum into two parts. The first one (which
we denote $\mathcal{S}_{\le}$) is taken over a cube $|q_\mu|\le Q$
except for $q=0$.
The rest is denoted $\mathcal{S}_{>}$. 

We estimate $\mathcal{S}_{>}$ first. We would like to show that as $t\to 0$ 
this sum vanishes faster than $t^{p-\frac n2}$ for arbitrary 
positive $p$ (for convenience $\frac n2$ is extracted explicitly).
Since $l$ and $r$ are smooth, their Fourier coefficients
are of rapid decay. This means that the partial sums of
$|l_{q} r_{-q}|$ outside of the cube vanish faster than
any power of $Q$, i.e., if $Q$ is sufficiently large,
\begin{equation}
\sum_{|q_\mu |> Q}|l_{q} \, r_{-q}| \le c_3 \,Q^{-m} \label{1319}
\end{equation}
for any chosen $m$. For reasons which will become clear later,
we choose $Q=c_4 \,t^{-1/(4(1+\beta))}$, $m=4(1+\beta) p +1$. The
inequality
(\ref{1319}) holds for a sufficiently large $c_4$ and a sufficiently
small $t$. Now we estimate
\begin{equation*}
|\mathcal{S}_{>}| \, t^{-p + \frac n2} \le 
c_4\,(4\pi)^{-\frac n2} \, t^{-p} \,Q^{-m} = c_5 \,t^{\frac
1{4(1+\beta)}}
\end{equation*}
so that the left hand side vanishes for any $p$ as $t\to 0$.

We now turn to $\mathcal{S}_{\le}$. This is a finite sum with at most
$Q^{n}$ terms. The Fourier coefficients entering this
sum are bounded by a constant, $|l_{-q}  r_q|\le c_6$. We also
have
\begin{equation*}
 -\frac{C^2}{4 |q|^{2(1+\beta)} t} \le
 -\frac{C^2}{4 |Q|^{2(1+\beta)} t} =
 - \frac{c_7}{t^{1/2} }\,.
\end{equation*}
Therefore,
\begin{equation*}
|\mathcal{S}_{\le}| \, t^{-p + \frac n2} \le 
c_6 \, (4\pi)^{-\frac n2} \, t^{-p} \, (2Q)^n 
e^ {- c_7/t^{1/2}}.
\end{equation*}
This expression vanishes at $t\to 0$ because of the exponential
damping. This completes the proof of (\ref{asym1}).

The proof of (\ref{asym2}) goes in the same way. The main observation
is that the sum over $\mathcal{K} \setminus \{0\}$ remains
exponentially
small, while the ``main'' terms (produced by $q\in \mathcal{Z}$) are
zero
since the first derivative of the geodesic distance vanishes in the
coincidence
limit.

\section{Beyond the Diophantine condition}

This section is an attempt to understand what happens if $\Theta$
is `in between' rational numbers and ``Diophantine numbers''. 
Consider the simplest case: $\T^2$ with 
\begin{equation*}
\Theta^{\mu\nu}=\theta 
\left( \begin{array}{cc} 0 & -1 \\ 1 & 0 \end{array} \right),
\end{equation*}
and $g={\rm diag} (1,1)$. 

To proceed, we need some results from number theory \cite{Bugeaud}.
Let $f\,:\, \R_{\geq 1} \rightarrow \R_{>0}$ be a continuous
function such that $x \rightarrow x^2\,f(x)$ is
non-increasing. Consider the set
$$
\mathcal{F}(f):=\set{\theta \in \R \, : \, \vert\theta q -p
\vert < qf(q) \text{ for infinitely many rational numbers
$\tfrac pq$} }
$$
The elements of $\mathcal{F}(f)$ are termed $f$-approximable. Note that
we cannot expect the above estimate to be valid for all rational 
numbers $\tfrac pq$ since for all irrational numbers $\th$, the set of
fractional values of $(\th q)_{q\geq 1}$ is dense in $[0,1]$.

Then, there exists an uncountable set of
real numbers $\theta/(2\pi)$ which are $f$-approximable but not
$cf$-approximable for any $0<c<1$, see \cite[Exercise 1.5]{Bugeaud}.

Let us choose 
\begin{equation}
f(x)= (2\pi x)^{-1}e^{-c_{8}x},\label{fx}
\end{equation}
$c_{8}>1$, and fix a constant
$c_{9} < 1$. Let us pick a $\theta$ which is $f$-approximable,
but not $c_{9}f$-approxi\-ma\-ble. We restrict our attention to the 
functions $l$ and $r$ which do not depend on the second coordinate
$x^2$
and $l_qr_{-q}=l_{-q}r_q$.
Then, according to (\ref{1847}),
\begin{equation*}
 \Tr \left( L(l) R(r) e^{-t\Delta}\right) = \frac 1{4\pi t}
\big( l_0 \, r_0
+ 2 \sum_{q_1 > 1}  l_{q_1,0}\, r_{-q_1,0} 
\sum_{k_2 \in \mathbb{Z}} e^{-(\theta q_1 -2\pi k_2)^2/4t}\big)
\end{equation*}
up to exponentially small terms. The first term in parentheses is
our standard result. Let us consider the ``correction'' term only.
Let $k_2^{(0)}(q_1)$ be an integer which minimizes the distance
to $\theta q_1/2\pi$. The sum over $k_2 \ne  k_2^{(0)}(q_1)$ is
exponentially small. Consequently, the correction term becomes
\begin{equation}
T(t):= \frac 1{2\pi t} \sum_{q_1 > 1} l_{q_1,0}\, r_{-q_1,0} \,\,
e^{ -(\theta q_1 +2\pi k_2^{(0)}(q_1))^2/4t } + \mbox{ e.s.t.}
\label{1816}
\end{equation}
For a Diophantine $\theta$, the whole correction term (\ref{1816})
is exponentially small. For $\theta/(2\pi)\in \mathbb{Q}$ this term
in $\mathcal{O}(1/t)$. Below we work out two explicit examples
and show that for the values of $\theta$ which we consider in this
section the correction term is, in general, neither one nor the other.

\quad

\noindent {\bf Example 1}.
Let us take 
\begin{equation*}
l_{q_1,0}\, r_{-q_1,0}=e^{-\alpha | q_1|}.
\end{equation*}
According to our assumption, $\theta /(2\pi)$ is not 
$c_{9}f$-approximable. Consequently, for all but a finite
number of $q_1\in \mathbb{N}$, 
$|\theta q_1 -2\pi k_2^{(0)}(q_1)| >c_{9} \, e^{-c_{8} \, q_1}.$
Then we can estimate (\ref{1816}) as
\begin{equation*}
T(t)\le  \frac 1{2\pi t} \sum_{q_1=0}^\infty  e^{-\alpha |q_1|}
\exp \left(
-\frac {c_{9}^2 \, e^{-2c_{8} \,q_1}}{4t} \right)+ \mbox{ e.s.t.}
\end{equation*}
(adding or removing any finite number of terms in
this sum does not change it up to e.s.t.).
Now we use the Euler--Maclaurin formula to transform this sum
to an integral (with exponentially small correction terms):
\begin{equation*}
 \frac 1{2\pi t} \int_0^\infty dq \, e^{-\alpha q} \,
e^{-\frac {c_{9}^2 \, e^{-2c_{8} \, q}}{4t} }
\end{equation*}
This integral can be easily evaluated:
\begin{equation}
\label{est1}
T(t)\le (2\pi c_{8})^{-1} \Gamma ( \tfrac {\alpha}{2c_{8}})
c_{9}^{-\frac \alpha{c_{8}}} \cdot
t^{-1 +\frac{\alpha}{2c_{8}}} + \mbox{ e.s.t.}
\end{equation}

\quad

\noindent {\bf Example 2}. In our second example we shall obtain a
lower
bound on $T(t)$, though for a different choice of $l$ and $r$. 
According to our assumption, $\theta/(2\pi)$ is $f$-approximable with
$f$ given by (\ref{fx}). Therefore, for an infinite set of $q_j$,
$j=1,2,\dots$
\begin{equation}
0< \vert \theta q_j - 2\pi k^{(0)}(q_j) \vert < e^{-c_{8} q_j}
\label{1835}
\end{equation}
We suppose that the $q_j$ are ordered, $q_j < q_{j+1}$. Consequently
$j\le q_j$ and (\ref{1835}) yields $$
\vert \theta q_j - 2\pi k^{(0)}(q_j) \vert < e^{-c_{8} j}.
$$
Let us now take
\begin{equation*}
l_{q,0}\, r_{-q,0}=\delta_{q,q_{j}}\,e^{-\alpha j}.
\end{equation*}
By repeating the 
arguments from previous example, one obtains
\begin{equation}
\label{est2}
T(t)\ge (2\pi c_{8})^{-1} \Gamma ( \tfrac{\alpha}{2c_{8}})
 \cdot
t^{-1 +\frac{\alpha}{2c_{8}}} + \mbox{ e.s.t.}
\end{equation}

\quad

The two estimates (\ref{est1}), (\ref{est2}) suggest 
that for non-Diophantine irrational
$\theta/(2\pi)$ one has to expect power-law corrections to the 
asymptotics (\ref{asym1}). These power-law corrections are unstable
in the sense that they crucially depend on the asymptotic behavior 
of $l_q\,r_{-q}$ for large $q$.

\subsection*{Acknowledgments}

\hspace{\parindent}
We thank Yann Bugeaud and Christian Mauduit for their help with 
number theory and Harald Grosse for helpful comments and discussions.
We are indebted to Jos\'e Gracia--Bond\'ia and Joseph
V\'arilly for philological advices and constructive comments.
The work of DV was supported in part by the DFG project BO 1112/13-1
(Germany) and by the grant RNP 2.1.1.1112 (Russia).

\end{document}